\documentclass{aa}  

\usepackage{graphicx}
\usepackage{txfonts}
\usepackage{hyperref}
\hypersetup{colorlinks=true,linkcolor=blue,citecolor=blue,filecolor=blue,urlcolor=blue}

\newcommand{\tabhead}[1]{\textbf{#1}}

\newcommand{\comment}[1]{}

\usepackage[flushleft]{threeparttable}

\newcommand{\targ}{YY\,Dra}
\newcommand{\kms}{ km~s$^{-1}$}
\usepackage{tablefootnote}

\begin{document}

   \title{The mass of the white dwarf in YY Dra (=DO~Dra): Dynamical measurement and comparative study with X-ray estimates}

   \subtitle{}

   \author{A. \'Alvarez-Hern\'andez
          \inst{1,2}
          \and
          M. A. P. Torres
          \inst{1,2}
          \and
          T. Shahbaz
          \inst{1,2}
          \and
          P. Rodr\'\i guez-Gil
          \inst{1,2}
          \and
          K. D. Gazeas
          \inst{3}
          \and \\
          J. S\'anchez-Sierras
          \inst{1,2}
          \and
          P. G. Jonker
          \inst{4}
          \and
          J. M. Corral-Santana
          \inst{5}
          \and
          J. A. Acosta-Pulido
          \inst{1,2}
          \and
          P. Hakala
          \inst{6}
          }

   \institute{Instituto de Astrof\'\i sica de Canarias, E-38205 La Laguna, Tenerife, Spain\
         \and
             Departamento de Astrof\'\i sica, Universidad de La Laguna, E-38206 La Laguna, Tenerife, Spain\
             \and
             Section of Astrophysics, Astronomy and Mechanics, Department of Physics, National and Kapodistrian University of Athens, GR-15784 Zografos, Athens, Greece\
             \and
             Department of Astrophysics/IMAPP, Radboud University Nijmegen, P.O.~Box 9010, 6500 GL Nijmegen, The Netherlands\
             \and
             European Southern Observatory, Alonso de C\'ordova 3107, Casilla 19001, Vitacura, Santiago, Chile\
             \and
             Finnish Centre for Astronomy with ESO (FINCA), Quantum, Vesilinnantie 5, FI-20014 University of Turku, Finland\
             }


 
\abstract
{We present a dynamical study of the intermediate polar cataclysmic variable \targ\ based on time-series observations in the $K$ band, where the donor star is known to be the major flux contributor. We covered the $3.97$-h orbital cycle with 44 spectra taken between $2020$ and $2022$ and two epochs of photometry observed in 2021 March and May. One of the light curves was simultaneously obtained with spectroscopy to better account for the effects of irradiation of the donor star and the presence of accretion light. From the spectroscopy, we derived the radial velocity curve of the donor star metallic absorption lines, constrained its spectral type to M0.5--M3.5 with no measurable changes in the effective temperature between the irradiated and non-irradiated hemispheres of the star, and measured its projected rotational velocity $v_\mathrm{rot} \sin  i  = 103 \pm 2 \, \mathrm{km}\,\mathrm{s}^{-1}$. Through simultaneous modelling of the radial velocity and light curves, we derived values for the radial velocity semi-amplitude of the donor star, $K_2 = 188^{+1}_{-2}$~\kms , the donor to white dwarf mass ratio, $q=M_2/M_1 = 0.62 \pm 0.02$, and the orbital inclination, $i={42^{\circ}}^{+2^{\circ}}_{-1^{\circ}}$. These binary parameters yield dynamical masses of $M_{1} = 0.99^{+0.10}_{-0.09} \, \mathrm{M}_{\odot}$ and $M_2 = 0.62^{+0.07}_{-0.06} \, \mathrm{M}_{\odot}$ ($68$ per cent confidence level). As found for the intermediate polars GK Per and XY Ari, the white dwarf dynamical mass in YY~Dra significantly differs from several estimates obtained by modelling the X-ray spectral continuum. 
}

   \keywords{accretion, accretion discs -- binaries: close -- novae, cataclysmic variables -- stars: individual: YY~Dra
               }

\titlerunning{Stellar masses in YY~Dra}
   \maketitle
%

\section{Introduction} \label{sec-intro}

Intermediate polars (IPs) are a subgroup of cataclysmic variables (CVs) consisting of a non-degenerate donor star that fills its Roche lobe and transfers mass to a moderately magnetic ($\sim 10^4-10^6$~G) white dwarf (WD) primary star (see \citealt{patterson-94} for a review). In contrast to polars \citep{chanmugam-77,cropper-90}, IPs usually have an accretion disc. However, it is truncated in the close proximity of the WD, where the magnetic field takes control of the motion of the plasma.

\targ\ (also known as DO~Dra\footnote{A summary of the controversy regarding the name of YY~Dra is available in \textit{The Catalog of IPs and IP Candidates} by Koji Mukai: \url{https://asd.gsfc.nasa.gov/Koji.Mukai/iphome/systems/yydra.html}}) is an IP which was identified as the counterpart of the hard X-ray source 3A~1148~+~719 by \cite{patterson-82}. \cite{williams-83} derived an orbital period of approximately four hours by fitting emission-line radial velocity curves obtained from optical spectroscopic data. The WD has two nearly symmetric magnetic poles with similar emission properties, and it rotates with a spin period of $\simeq 529$~s, causing the ultraviolet (UV) and X-ray flux to be modulated as a double pulse \citep{patterson-92,haswell-97,szkody-2002}. YY~Dra also exhibits significant brightness variability, from dwarf nova outbursts \citep{simon-2000,szkody-2002,andronov-2018} to low states associated with significant reductions in accretion \citep{covington-22,katherine-22}.

\cite{friend-90} obtained $I$-band spectroscopy of \targ\ with $\simeq 80$~\kms\ resolution and, by comparing their average spectrum with spectral templates, they supported an M3~V donor star with a rotational broadening $v_\mathrm{rot} \sin  i~=~110~\pm~10$~\kms . They also obtained a radial velocity curve with full orbit coverage by cross-correlating the individual spectra against an M3~V template in the region of the Na~\textsc{i} absorption doublet at $\simeq 8190$~\AA. According to the authors, this radial velocity curve does not show significant eccentricity. Therefore, they fitted a circular orbit, resulting in a velocity semi-amplitude $K_{\mathrm{Na \textsc{i}}} = 202 \pm 3$~\kms . \cite{mateo-91} also took $I$-band time-resolved spectroscopy ($\simeq 65$~\kms\ resolution) and suggested an M4 $\pm 1$ spectral type for the donor star based on visual comparison of the TiO bands in the average YY~Dra spectrum with those in M-dwarf spectral templates. They analysed the Na~\textsc{i} absorption doublet radial velocity curve, obtaining from a sine fit $K_{\mathrm{Na \textsc{i}}} = 193 \pm 8$~\kms . The curve shows appreciable deviations from the fit at orbit quadratures with an apparent eccentricity $e = 0.056 \pm 0.026$. This, together with the orbital dependence found in the strength of the narrow Ca~\textsc{ii} triplet emission line components, was identified as a clear sign of UV/X-ray heating of the donor star. Utilising $K_{\mathrm{Na \textsc{i}}}$, the apparent eccentricity and the irradiation models for the Na~\textsc{i} doublet by \cite{marsh-88}, they derived the radial velocity semi-amplitude for the centre of mass of the donor star, $K_2 = 184 \pm 10$~\kms . Later, \cite{haswell-97} used $I$-band photometry and 14-yr time baseline H$\alpha$ spectroscopy, along with the previously published Ca~\textsc{ii} radial velocity measurements from \cite{mateo-91}, to better constrain the orbital period to $P = 0.16537398(17)$~d. \cite{joshi-2012} further refined it to $P = 0.16537424(2)$~d using long-term, low-cadence $H$- and $K$-band photometry taken between 2007 and 2009. This was achieved by comparing the phase shifts between the observed and calculated times for the maximum approach of the donor star employing the ephemeris given by \cite{haswell-97}. Finally, \cite{katherine-22} noted that Joshi's orbital period results in a phase shift of $\simeq 0.02$ orbital cycle in the ellipsoidal modulation observed in the 2019--2020 Transiting Exoplanet Survey Satellite (TESS) light curve. They slightly adjusted Joshi's orbital period to eliminate this phase shift, establishing the most accurate value in the literature: $P = 0.16537420(2)$~d, equivalent to $3.9689808(5)$~h.

Other fundamental parameters of YY~Dra are poorly constrained. \cite{mateo-91} and \cite{haswell-97} derived the donor to white dwarf mass ratio, $q=M_2/M_1$, from the ratio between the radial velocity semi-amplitude of emission and absorption lines. They obtained $q=0.48 \pm 0.07$ and $q=0.45 \pm 0.05$, respectively. However, these values could suffer from inaccuracy, since emission lines in CVs typically do not trace the movement of the centre of mass of the WD (see e.g. \citealt{bitner-2007} and references therein). \cite{haswell-97} also derived $q=0.61 \pm 0.09$ from $K_{\mathrm{Na \textsc{i}}} = 202 \pm 3$~\kms\ and $v_\mathrm{rot} \sin  i = 110 \pm 10$~\kms\ measurements by \cite{friend-90}. \cite{mateo-91} and \cite{haswell-97} estimated the orbital inclination to be $i = 42^{\circ} \pm 5^{\circ}$ and $i = 45^{\circ} \pm 4^{\circ}$, respectively. In both calculations they assumed that the donor star follows the mass-radius relationship for unevolved late-type dwarfs and adopted the respective $q$ values. Later, \cite{joshi-2012} found $i = 41^{\circ} \pm 3^{\circ}$ by fitting his phase-folded, long-term $H$-band light curve with the Wilson-Devinney code \citep{wilson-2020}. This estimate from modelling the combined light curve is open to potential systematic errors due to the adoption of 3150~K for the donor star effective temperature and $q=0.45$ \citep{haswell-97}. It could also be affected by potential changes in the morphology of the YY Dra ellipsoidal light curve resulting from variations in the accretion component and UV/X-ray heating of the donor star. On the whole, these studies constrain the WD and donor star masses to $\simeq 0.6 - 1.0 \, \mathrm{M}_{\odot}$ and $\simeq 0.3 - 0.5 \, \mathrm{M}_{\odot}$ ($1 \sigma$), respectively. Estimates of the WD mass have also been obtained from X-ray data, with earlier studies resulting in a broad range of values ($\simeq 0.4 - 1.0 \, \mathrm{M}_{\odot}$ range at $1 \sigma$) and most recent ones pointing towards $\simeq 0.8 \, \mathrm{M}_{\odot}$ (see Sect.~\ref{subsec-m1x} for further details). 

In this article, we obtain dynamical masses for YY~Dra through partially simultaneous spectroscopy and photometry in the $K$ band, where the donor star has been found to be the dominant source of light \citep{mateo-91,harrison-16}. The paper is structured as follows: Sect.~\ref{sec-obs} presents our near-infrared spectroscopic and photometric observations, along with the data reduction process. In Sects.~\ref{sec-k2}, \ref{sec-spectralclassification} and~\ref{sec-rotationalbroadening}, we use the stellar metallic lines in the spectra to obtain their radial velocity curve, constrain the spectral type of the donor star, and measure the rotational broadening, respectively. Following this, Sect.~\ref{sec-modelling} describes the fitting of the radial velocity curve and the photometry with a model that includes the effects of irradiation of the donor star to constrain all orbital parameters, ultimately allowing us to derive the dynamical masses. Sect.~\ref{sec-discuss} is dedicated to discussing our main results and comparing our WD mass with previous estimates from X-ray studies. Finally, Sect.~\ref{sec-conclusions} provides our conclusions.

\section{Observations and data reduction}\label{sec-obs}

\subsection{GTC/EMIR spectroscopy}\label{subsec-spectroscopy}

\begin{table*}
\caption[]{Log of the \targ\ GTC/EMIR spectroscopy.}
\label{tab:observaciones_espectroscopia}
\centering
\begin{tabular}{l c c c c c c}
\hline\noalign{\smallskip}
Date  & Epoch & ABBA & Orbital phase & Airmass & Seeing & $K$\\
\textbf{} &  & cycles & coverage & \textbf{} & (arcsec) & (mag) \\
\hline\noalign{\smallskip}
2020 February 13 & 1 & 12 & $0.17 - 0.59$ & $1.36 - 1.38$ & $0.76 - 0.89$ & 12.2\\ 
2021 May 5 & 2 & 12 & $0.88 - 0.30$ & $1.36 - 1.40$ & $0.67 - 0.94$ & $12.2-12.4$\\ 
2021 May 5 & 2 & 4 & $0.75 - 0.85$ & $1.55 - 1.62$ & $1.47 - 1.57$ & 12.4\\ 
2022 May 12 & 3 & 8 & $0.48 - 0.73$  & $1.37 - 1.42$ & $0.86 - 0.94$ & 12.3\\ 
2022 May 17 & 3 & 8 & $0.97 - 0.31$ & $1.44 - 1.61$ & $0.67 - 0.83$ & $12.4-12.5$\\ 
\hline\noalign{\smallskip}
\end{tabular}
\tablefoot{The orbital phases were calculated using the ephemeris given by \cite{katherine-22}. Each ABBA nodding cycle had a 480~s on-source time. We measured the seeing as the FWHM of the spatial profile around $\lambda = 2.2~\mu\mathrm{m}$. Differential photometry of the target relative to the nearby field star YY Dra–8 of \cite{hh-95} in the acquisition images yielded the magnitudes. TYC 4395-97-1, the star used in Sect.~\ref{subsec-photometry}, was out of the linear range of the detector.}
\end{table*}

Time-resolved near-infrared spectroscopy of \targ\ was obtained using the EMIR spectrograph \citep{garzon-16, garzon-22} on the 10.4-m Gran Telescopio Canarias at the Observatorio del Roque de los Muchachos on the island of La Palma, Spain. The target was observed on four different nights in the $2020 - 2022$ period in queue mode, using time constraints and without observing telluric standards, in a best effort to properly sample most of its 3.97-h orbit. The Configurable Slit Unit was set up to create a 0.6-arcsec wide long slit positioned one arcmin to the left of the field of view centre. This, combined with the K grism, provided coverage of the $2.08-2.43~\mu\mathrm{m}$ wavelength range with a dispersion of 1.71~\AA\,$\mathrm{pixel}^{-1}$ and a resolution of $\simeq 4.8$~\AA\ full-width at half-maximum (FWHM), which was measured using the atmospheric OH emission lines. This translates into $\simeq 65$~\kms\ resolution at $2.2~\mu\mathrm{m}$. 

Each observing visit/block to the target consisted of four consecutive ABBA nodding cycles with 12~arcsec offsets and individual exposures of 120~s. In total, we obtained 44 ABBA cycles on four different nights (see Table~\ref{tab:observaciones_espectroscopia} for a log of the observations). In addition, we observed the spectral type template stars Gl~176 (M2~V), Gl~797~B (M2~V) and LHS~3558 (M3~V) with the same instrument setup as employed for YY~Dra. They have low rotational velocities and their effective temperatures \citep{rojas-ayala-2012} are within our spectroscopic constraints for the donor star in YY~Dra (Sect.~\ref{sec-spectralclassification}). We assessed the image quality of the YY~Dra and template star spectra by performing Gaussian fitting on the spatial profiles. Our findings indicate that all spectra were taken under slit-limited conditions (see Table~\ref{tab:observaciones_espectroscopia}).

For data reduction we used version 0.17.0 of \textsc{pyemir} \citep{cardiel-19}. After implementing dark and flat-field corrections, rectification of geometric distortions and wavelength calibration were performed in two stages. Initially, a preliminary calibration, computed by the instrument team, was applied to the 2D frames. Subsequently, refinement of this calibration was performed using atmospheric OH emission lines, allowing for more precise delineation of slitlet boundaries and enhancing wavelength calibration for each (refer to \citealt{cardiel-19} and the \textsc{pyemir} manual\footnote{\url{https://guaix-ucm.github.io/pyemir-tutorials/}} for additional details). In all cases, the root mean square scatter of the fits was $0.5-0.8$~\AA\ (equivalent to $7-11$~\kms\ at $2.2~\mu\mathrm{m}$). We established the stability of the wavelength calibration by examining the radial velocities of the star TYC 4395-97-1, which was positioned within the slit during each observation of YY~Dra. The calibration, extraction and telluric correction of these spectra were performed as described for YY~Dra. We found a standard deviation of $2.4$~\kms . The four spectra of each nodding cycle were averaged, sky subtracted and dithering corrected. The extraction of the resulting average spectra was performed with the \textit{apall} task in \textsc{iraf}\footnote{{\sc iraf} is distributed by the National Optical Astronomy Observatories.}.

To remove the telluric absorption features, we derived the atmospheric transmission spectrum by fitting the telluric absorptions in the science and M-dwarf template spectra using version 1.5.9  of \textsc{molecfit} \citep{smette-15}. As for the dynamical study of XY Ari \citep{ayoze-2023}, we utilised custom \textsc{python} scripts to adapt the EMIR data format for use in \textsc{molecfit}. During the fitting process, we excluded wavelength regions expected to contain features characteristic of M-type dwarfs. 

Prior to the analysis, we imported the telluric-corrected spectra into \textsc{molly} and corrected them for the Earth motion, shifting the spectra to the heliocentric rest frame. For each spectrum, the time corresponds to the middle of the effective exposure and is expressed in heliocentric Julian days (UTC). The spectra were normalised using a spline-fitted continuum and then rebinned into a logarithmic scale that is uniform in velocity. In the same way, we imported to \textsc{molly}, normalised and rebinned the high signal-to-noise ratio (S/N $\geq 100$) public library spectra presented in Sect.~\ref{sec-spectralclassification}, which were used for the spectral classification of the donor star.

\subsection{NOT/NOTCAM $K_{\mathrm{s}}$-band photometry}\label{subsec-photometry}

\comment{
\begin{table}
\caption[]{Log of the time-resolved $K_{\mathrm{s}}$-band photometry.}
\setlength{\tabcolsep}{0.9ex}
\label{tab:observaciones_fotometria}
\begin{center}
\begin{tabular}{l c c c}
\hline\noalign{\smallskip}
Date &\tabhead{$\#$} &\tabhead{$T_\mathrm{exp}$} & Coverage\\
 &\tabhead{} &(s) & (h)\\
\hline\noalign{\smallskip}
2021 March 31 & 596 & 7.2 & 5.82 \\
2021 May 5 & 375 & 7.2 & 4.20 \\
\hline\noalign{\smallskip}
\end{tabular}
\end{center}
\end{table}
}

Time-resolved $K_{\mathrm{s}}$-band photometry of \targ\ was obtained using the Nordic Optical Telescope near-infrared Camera and spectrograph (NOTCAM) mounted on the 2.56-m Nordic Optical Telescope (NOT) at the Roque de los Muchachos. This instrument has a $1024 \times 1024$ pixel MCT detector divided into four $512 \times 512$ pixel quadrants. We employed the Wide-Field mode, which provides a field of view of $4 \times 4$ arcmin with a plate scale of 0.235 $\mathrm{arcsec} \, \mathrm{pixel}^{-1}$. The observing strategy was a  ``5-point dice'' dithering pattern with $7.2$-s exposure time at each dither position. The target was observed on 2021 March 31 and May 5, covering $\simeq1.4$ and $\simeq1.0$ orbital cycles, respectively. The time-resolution of our observations was $\simeq 27$ seconds on 2021 March 31 and $\simeq 40$ seconds on 2021 May 5. This arised from the different read out modes we used for each observation (ramp-sampling versus reset-read-read, see the NOTCAM manual for further details). The average seeing was approximately 0.9 arcsec on the first night and approximately 0.8 arcsec on the second night. The observations on 2021 May 5 were simultaneous with GTC/EMIR spectroscopy.

Non-linearity correction, flat-field correction, sky-subtraction, and alignment of all the images were carried out using version 2.6 of the NOTCAM \textsc{iraf} reduction package\footnote{\url{https://www.not.iac.es/instruments/notcam/quicklook.README}}. Differential photometry of \targ\ with a variable aperture was performed in each individual image relative to the field star TYC 4395-97-1 ($K_{\mathrm{s}} = 10.51 \pm 0.016$~mag, \citealt{2mass}) using version 1.2.0 of the HiPERCAM pipeline\footnote{\url{https://github.com/HiPERCAM}}. We used YY Dra--8 \citep{hh-95} as the check star, for which we obtained a mean magnitude and standard deviation of $K_{\mathrm{s}} = 12.16 \pm 0.02$~mag from the two nights of observations. For the subsequent analysis, we excluded all photometry in which the YY~Dra--8 magnitude deviated by more than $2 \, \sigma$ ($=0.04$~mag) from the mean. We also neglected data points with an uncertainty in the target's magnitude greater than $0.03$~mag. These two criteria resulted in the exclusion of 26 data points (out of a total of 596) and 40 (out of a total of 375) in the first and second light curves, respectively.

\begin{figure}
\centering \includegraphics[height=7.5cm]{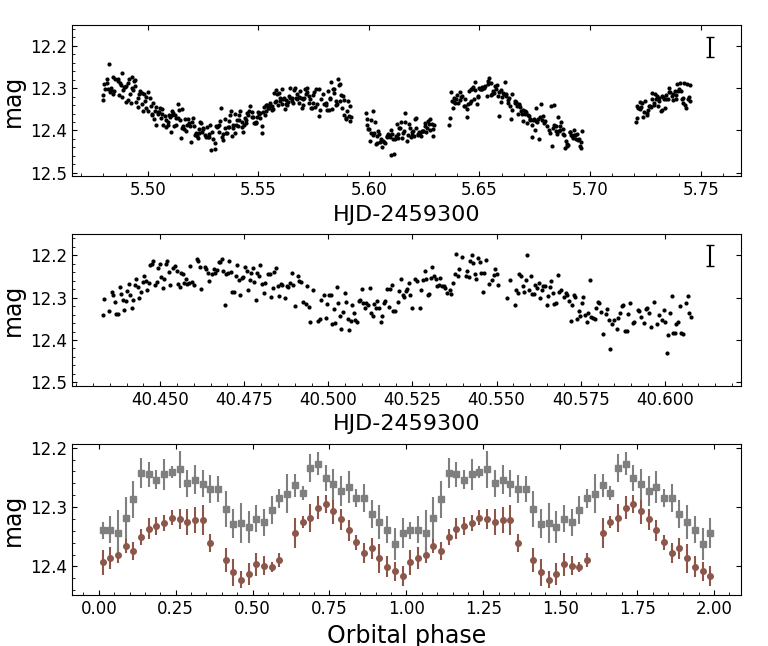}
\caption{\label{fig:lightcurves} $K_{\mathrm{s}}$-band light curves derived from the NOT/NOTCAM photometry taken at 2021 March 31 (top panel) and 2021 May 5 (middle panel). Error bars on the right indicate the typical uncertainty, which is $\simeq 0.025$~mag in both cases. The bottom panel shows both light curves, phase-folded and phase-binned. Brown circles correspond to March 31, while grey squares represent May 5. To enhance clarity, the orbital cycle is repeated.}
\end{figure}

The top and middle panels in Fig.~\ref{fig:lightcurves} show the resulting light curves, which display a $\simeq0.1$~mag peak-to-peak ellipsoidal modulation with superimposed lower amplitude flickering variability. In the bottom panel of Fig.~\ref{fig:lightcurves} both light curves are phase-folded and averaged into $40$ orbital phase bins using the ephemeris from \cite{katherine-22}: $P = 0.16537420$~d, $T_0 = 2446863.4376$~(HJD). The value and the uncertainty of each bin corresponds to the mean and the standard deviation of the data points within the bin. Similarities and changes in the morphology of the ellipsoidal modulation become apparent in the resulting light curves. The source was slightly brighter on 2021 May 5. Specifically, the mean magnitude and standard deviation for the May 5 light curve are $K_{\mathrm{s}} = 12.29 \pm 0.04$~mag, compared to $12.36 \pm 0.04$~mag for March 31. The two maxima of each phased light curve are at the same flux level (within the errors), indicating the absence of disc or star spots with significant flux contribution.

\begin{table*}
\caption[]{Best-fit parameters of the circular and elliptical fits to the radial velocity measurements.}
\centering
\begin{tabular}{lcccc}
\hline\noalign{\smallskip}
\noalign{\smallskip}
\noalign{\smallskip}
\noalign{\smallskip}
\multicolumn{4}{c}{\textbf{Circular fits: $V (t) = \gamma + K_{\mathrm{abs}} \, \mathrm{sin} \, \left[ \frac{2 \pi}{P}(t-T_{0}) \right]$}}\\
\\
Data set & $\gamma$ & $K_{\mathrm{abs}}$ & $\chi^{2}/$ & dof\\
& (\kms) & (\kms) & dof & \\
\hline\noalign{\smallskip}
2020 February & $-3$(3) & 181(3) & 2.52 & 10\\
2021 May & $-6$(3) & 186(2) & 0.72 & 14\\
2022 May & $-3$(3) & 184(3) & 1.17 & 14\\
2021 May + 2022 May & $-5$(3) & 185(1) & 0.94 & 30\\
\hline\noalign{\smallskip}
\noalign{\smallskip}
\noalign{\smallskip}
\noalign{\smallskip}
\multicolumn{4}{c}{\textbf{Elliptical fits:} $V (t) = \gamma + S_1 \, \mathrm{sin} \, x + C_1 \, \mathrm{cos} \, x + S_2 \, \mathrm{sin} \, 2x + C_2 \, \mathrm{cos} \, 2x$}\\
\\
Data set & $\gamma$ & $e = \sqrt{\frac{S_2^2+C_2^2}{S_1^2+C_1^2}}$ & $\chi^{2}$/ & dof\\
& (\kms) & & dof &\\
\hline\noalign{\smallskip}
2020 February & 5(20) & 0.08(8) & 1.42 & 7 \\
2021 May & $-$2(6) & 0.02(2) & 0.62 & 11 \\
2022 May & $-5$(3) & 0.03(2) & 0.95 & 11\\
2021 May + 2022 May & $-6$(3) & 0.03(1) & 0.78 & 27\\
\hline\noalign{\smallskip}
\end{tabular}
\label{tab:rvcurve_params}
\tablefoot{Numbers in brackets specify the uncertainty in the last digit, where we have quadratically added the 2.4~\kms\ standard deviation to the $\gamma$ errors to account for the uncertainty in the measurement of the intrinsic radial velocity of the template.}
\end{table*}

\begin{figure*}
\centering \includegraphics[height=13.5cm]{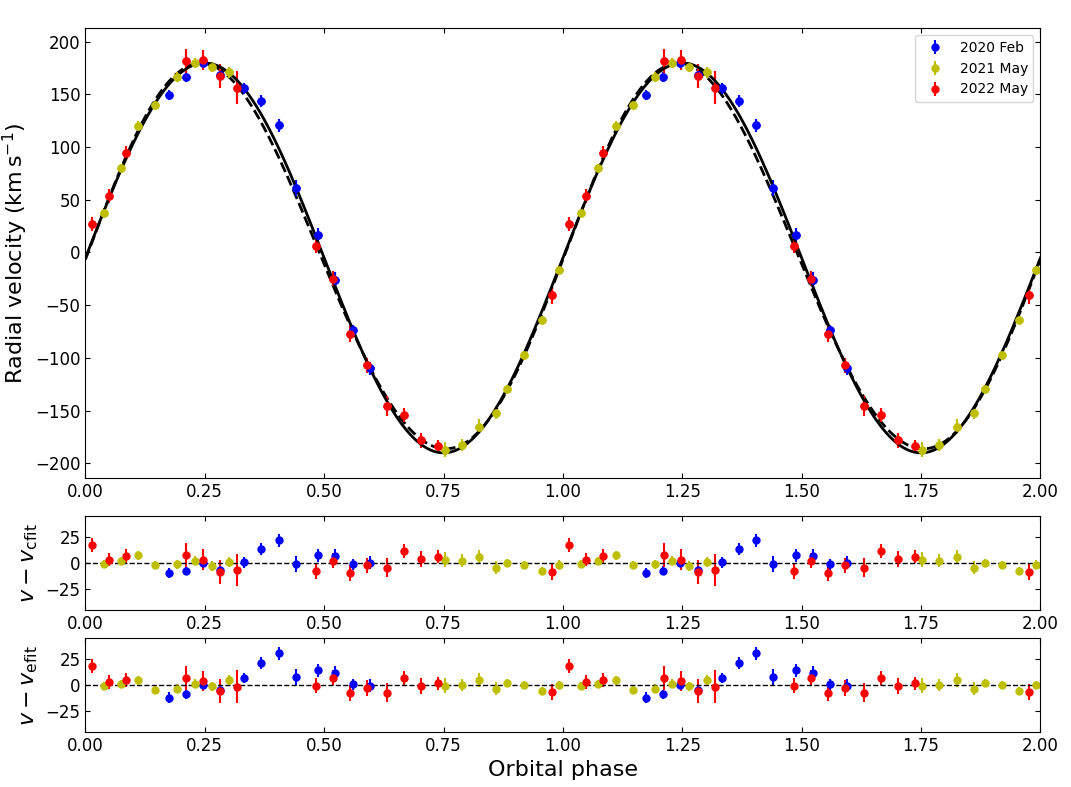}
\caption{\label{fig:radialvelocitycurve} Phase-folded heliocentric radial velocity curve of the donor star in \targ , obtained by cross-correlating our individual spectra of the target with the M2~V template spectrum Gl~797~B. The top panel shows the measured radial velocities along with the best circular (solid line) and elliptical (dashed line) fits. The middle panel displays the residuals from the circular fit, while the bottom panel represents those from the elliptical fit. The orbital cycle is shown twice for the sake of clarity. We excluded the 2020 February radial velocities from the fitting process. Data points from this epoch at phases $0.17-0.21$ and $0.37-0.41$ clearly exhibit different behaviour.}
\end{figure*}

\section{Analysis and results}
Multiple metallic absorption lines arising from the photosphere of the donor star are clearly visible in the $K$-band spectral range \citep{harrison-16}. The most prominent lines are the Na~\textsc{i} doublet at $\simeq 2.21~\mu\mathrm{m}$ and the Ca~\textsc{i} doublet at $\simeq 2.26~\mu\mathrm{m}$, but there also some weaker features, primarily from Ti~\textsc{i} and Si~\textsc{i}. All the spectroscopic measurements presented in this article were obtained in the wavelength ranges $2.09 - 2.11~$, $2.12 - 2.15~$ and $2.18 - 2.27$~$\mu \mathrm{m}$, which contain absorption features free from contamination from emission lines. Wavelengths $>2.27$~$\mu \mathrm{m}$ were excluded from the analysis due to a significant drop in S/N. In the remainder of this paper, uncertainties are specified at the $68$ per cent confidence level.

\subsection{Radial velocity curve}\label{sec-k2}

We used the absorption lines of the donor star to derive its radial velocity curve by cross-correlating each YY~Dra spectrum with the M2~V template Gl~797~B (see Sects.~\ref{sec-spectralclassification} and~\ref{sec-rotationalbroadening}). Before cross-correlation, the template spectrum was corrected for its intrinsic radial velocity (determined by fitting the core of the Na~\textsc{i} doublet at $\simeq 2.21~\mu\mathrm{m}$ with a double Gaussian), broadened to match the projected rotational velocity of the donor star (measured in Sect.~\ref{sec-rotationalbroadening}), and rebinned onto the same logarithmic scale as the YY~Dra spectra. The 2.4~\kms\ uncertainty in the wavelength calibration (Sect.~\ref{sec-obs}) was quadratically added to the statistical error of each radial velocity measurement. As a validation, we repeated the measurements using the observed LHS 3558 template (M3~V) and found similar values within the errors for all the radial velocities.

We expect the photospheric lines to be affected by irradiation, with the impact on the radial velocities potentially changing from epoch to epoch due to long-term accretion variability. Consequently, we conducted analyses of our radial velocity curves for different epochs (see Table~\ref{tab:observaciones_espectroscopia}) with both circular and elliptical orbit fits. During this process, we treated all the data from 2022 May 12 and 17 as a single epoch due to the insufficient data points on each night to achieve successful fits separately. For the circular fit, we employed a sinusoidal equation for the radial velocity $V(t)$:
\begin{equation}
\label{eq:seno}
V (t) = \gamma + K_{\mathrm{abs}} \, \mathrm{sin} \, \left[ \frac{2 \pi}{P}(t-T_{0}) \right] ,
\end{equation}  

\noindent where $\gamma$ is  the  heliocentric systemic velocity, $K_{\mathrm{abs}}$ denotes the observed radial velocity semi-amplitude, $P$ indicates the orbital period and $T_0$ represents the time of closest approach of the donor star to the observer. For the elliptical fit, we used a second-order Fourier series of the form:
\begin{equation}
\label{eq:elliptical_fit}
V (t) = \gamma + S_1 \, \mathrm{sin} \, x + C_1 \, \mathrm{cos} \, x + S_2 \, \mathrm{sin} \, 2x + C_2 \, \mathrm{cos} \, 2x ,
\end{equation}  

\noindent where $x = 2 \pi \, (t-T_{0}) \, / \, P$ and the eccentricity is given by $e~=~\sqrt{(S_2^2+C_2^2) \, / \, (S_1^2+C_1^2)}$ for $e \leq 0.1$ (see e.g. \citealt{martin-89}).

In all fits, we fixed the orbital period and $T_0$ according to the ephemeris given by \cite{katherine-22}. Table~\ref{tab:rvcurve_params} presents the best-fit parameters. Both the 2021~May and 2022~May data sets are well-fitted with a circular orbit, with $\chi^2 / \mathrm{dof}$ close to 1 and $K_\mathrm{abs} \simeq 185$~\kms ,  where dof refers to the degrees of freedom of the fit. In addition, the resulting $e$ value in the elliptical fit is either consistent or very close to null. In contrast, the 2020 February data set is not well-represented by the circular fit ($\chi^2 / \mathrm{dof} > 2$), and the $e$ value from the elliptical fit appears larger (but consistent with $0$ within the errors). This difference may be explained by a different irradiation level during our 2020 February observations. In this regard, YY~Dra experienced a low state during which accretion greatly diminished approximately $5$ days before our 2020 February observations \citep{katherine-22} and the data from TESS show that it was increasing in brightness at the time of our spectroscopy.

We completed our analysis of the radial velocity curve by combining the 2021 May and 2022 May data sets, since they gave similar fit parameters within the errors when treated separately. Figure~\ref{fig:radialvelocitycurve} shows all the phase-folded radial velocities along with the best circular (solid line) and elliptical (dashed line) fits to the combined 2021~May~+~2022~May data points. The elliptical fit results in a small but significant eccentricity, $e = 0.03 \pm 0.01$. For this reason, the value of the radial velocity semi-amplitude of the donor star's centre of mass, $K_2$, will be presented in Sect.~\ref{sec-modelling}, where we concurrently model the 2021~May~+~2022~May radial velocities and the light curves obtained from our NOT/NOTCAM photometry (Sect.~\ref{subsec-photometry}). In this process, we take advantage of the partial simultaneity between the radial velocity measurements and the photometric data to model the irradiation.

\comment{
\begin{table}
\caption[]{Best-fit parameters of the circular fit to the radial velocity curve of the donor star. Numbers in brackets specify the uncertainty in the last digit. We quadratically added the 2.4~\kms\ uncertainty from the wavelength calibration to the errors of the $\gamma$ values. Degrees of freedom ($\mathrm{dof}) = 41$.}
\centering
\begin{tabular}{lcccccc}
\hline\noalign{\smallskip}
Template & Sp. & $\gamma$ & $K_{\mathrm{abs}}$ & $T_0$  & $\chi^{2}$/\\
& Type & (\kms) & (\kms) &  (HJD) & $\mathrm{dof}$\\
\hline\noalign{\smallskip}
Gl 797 B & M2~V & -4(3) & 184(1) & 2458893.5839(2) & 1.36 \\
LHS 3558 & M3~V & -6(3) & 185(1) & 2458893.5839(2) & 1.41 \\
\hline\noalign{\smallskip}
\end{tabular}
\label{tab:rvcurve_params}
\end{table}
}

\subsection{Spectral classification of the donor star}\label{sec-spectralclassification}

Previous classifications of the donor star in YY~Dra include: M3~V \citep{patterson-82}, M3--M5.5~V \citep{mukai-90}, M3~V \citep{friend-90} and M4~$\pm~1$~V \citep{mateo-91}. Additionally, \cite{harrison-16} finds the donor star to be consistent with a M1.8 dwarf with metallicity $\mathrm{[Fe/H]}~=~-0.20$~dex using a single $240$~s $K$-band spectrum with $\simeq 135$~\kms\ resolution at $2.2~\mu\mathrm{m}$. For this analysis, he employed empirical relationships of the equivalent width of absorption lines in M-dwarfs and comparison with synthetic templates. Taking advantage of our full orbital coverage, we decided to test these results by performing our classification of the donor star with two grids of spectral templates with different metallicities: one with $-0.30~<~\mathrm{[Fe/H]}~<~0.00$~dex and another one with $0.00~<~\mathrm{[Fe/H]}~<~0.30$~dex. We will refer to them as the subsolar and supersolar metallicity grids, respectively. They were constructed from publicly available high S/N spectra of M-type dwarf stars in \cite{rojas-ayala-2012}. Their spectral resolution is $R = \lambda / \Delta \lambda = 2700$, which equates to $\simeq 110$~\kms\ resolution at $2.2~\mu\mathrm{m}$, close to the total broadening in the YY~Dra data resulting from the convolution of the instrumental and rotational profiles (Sects.~\ref{subsec-spectroscopy} and ~\ref{sec-rotationalbroadening}). Tables~\ref{tab:templates_rojas_lowFe} and~\ref{tab:templates_rojas_highFe} list all the selected templates, providing their effective temperatures and [Fe/H] values.

\begin{table}
\caption[]{Spectral classification of the donor star in \targ\ using the grid of subsolar metallicity templates.}
\centering
\begin{tabular}{l c c c c}
\hline\noalign{\smallskip}
Template & Effective & [Fe/H] & $f$ & $\chi^{2}_{\mathrm{min}}/$\\ 
 & temperature & (dex) & & $\mathrm{dof}$ \\ 
  & (K) &  & & \\ 
\hline\noalign{\smallskip}
Gl 338 A & $4031 \pm 56$ & $-0.18 \pm 0.17$ & $0.74(4)$ & 12.67 \\
Gl 338 B & $3869 \pm 15$ & $-0.15 \pm 0.17$ & $0.74(5)$ & 12.51 \\
Gl 649 & $3733 \pm 20$ & $-0.04 \pm 0.17$ & $0.89(2)$ & 5.82 \\
Gl 686 & $3693 \pm 20$ & $-0.28 \pm 0.17$ & $1.29(2)$ & 4.14 \\
Gl 408 & $3526 \pm 18$ & $-0.09 \pm 0.17$ & $1.05(2)$ & 3.49 \\
LHS 3605 & $3437 \pm 16$ & $-0.28 \pm 0.17$ & $1.18(3)$ & 5.37 \\
Gl 687 & $3395 \pm 18$ & $-0.09 \pm 0.17$ & $1.05(2)$ & 4.59 \\
Gl 661 AB & $3272 \pm 28$ & $-0.31 \pm 0.17$ & $1.08(5)$ & 7.23 \\
LHS 1723 & $3054 \pm 69$ & $-0.06 \pm 0.17$ & $0.95(3)$ & 6.34 \\
GJ 1116 AB & $2896 \pm 18$ & $-0.12 \pm 0.17$ & $0.79(4)$ & 11.00 \\
\hline\noalign{\smallskip}
\end{tabular}
\label{tab:templates_rojas_lowFe}
\tablefoot{The number of dof is 819. Brackets mark the uncertainty in the last digit. The values for the effective temperature and [Fe/H] are from \cite{rojas-ayala-2012}.}
\end{table}

\begin{table}
\caption[]{Same as Table~\ref{tab:templates_rojas_lowFe}, but for the supersolar metallicity grid of templates.}
\centering
\begin{tabular}{l c c c c}
\hline\noalign{\smallskip}
Template & Effective & [Fe/H] & $f$ & $\chi^{2}_{\mathrm{min}}/$\\ 
 & temperature & (dex) & & $\mathrm{dof}$ \\ 
  & (K) &  & & \\ 
\hline\noalign{\smallskip}
Gl 212 & $3851 \pm 17$ & $0.03 \pm 0.17$ & $0.76(2)$ & 8.37 \\
HD 46375 B & $3663 \pm 15$ & $0.29 \pm 0.17$ & $0.73(2)$ & 6.75 \\
Gl 176 & $3581 \pm 20$ & $0.15 \pm 0.17$ & $0.83(2)$ & 4.83 \\
NLTT 14186 & $3530 \pm 18$ & $0.24 \pm 0.17$ & $0.80(1)$ & 4.20 \\
Gl 402 & $3334 \pm 23$ & $0.20 \pm 0.17$ & $0.87(2)$ & 4.34 \\
Gl 555 & $3288 \pm 27$ & $0.22 \pm 0.17$ & $0.88(2)$ & 4.37 \\
Gl 268 AB & $3184 \pm 40$ & $0.10 \pm 0.17$ & $0.90(2)$ & 5.31 \\
Gl 905 & $3058 \pm 65$ & $0.19 \pm 0.17$ & $0.82(2)$ & 7.19 \\
\hline\noalign{\smallskip}
\end{tabular}
\label{tab:templates_rojas_highFe}
\end{table}

\begin{figure*}
\sidecaption
\includegraphics[width=12cm]{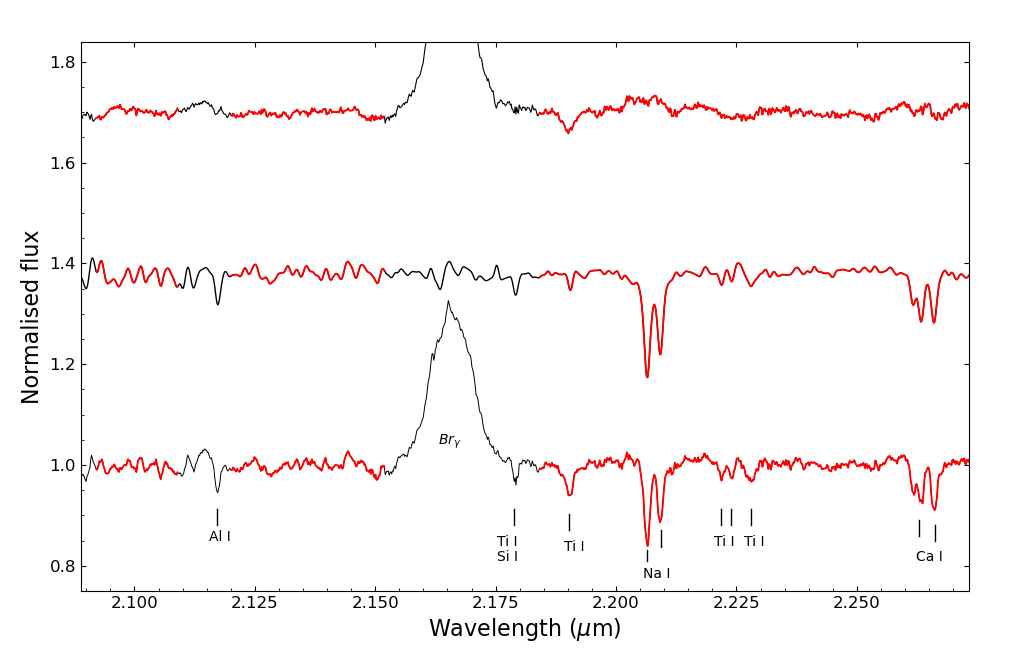}
\caption{\label{fig:optsub} Example of the application of the optimal subtraction technique. The order of the spectra, from bottom to top, is as follows: \targ\ average in the rest frame of the donor star, Gl~402 template from \cite{rojas-ayala-2012} broadened to match the YY~Dra average, and the residual after subtraction of the broadened and scaled template. Red sections in the spectra indicate the wavelength regions used for the analysis. For clarity, the template and residual spectra have been vertically shifted. We mark the most important absorption lines according to the NASA IRTF spectral library atlas \citep{rayner-2009}. Optimal subtraction removes all lines of the donor star, except for the Ti~\textsc{i} line at $2.1903 \, \mu \mathrm{m}$. This is also true for the other templates and in \cite{harrison-16}, indicating that this difference is intrinsic to YY~Dra.}
\end{figure*}

\begin{figure}
\centering \includegraphics[height=9cm]{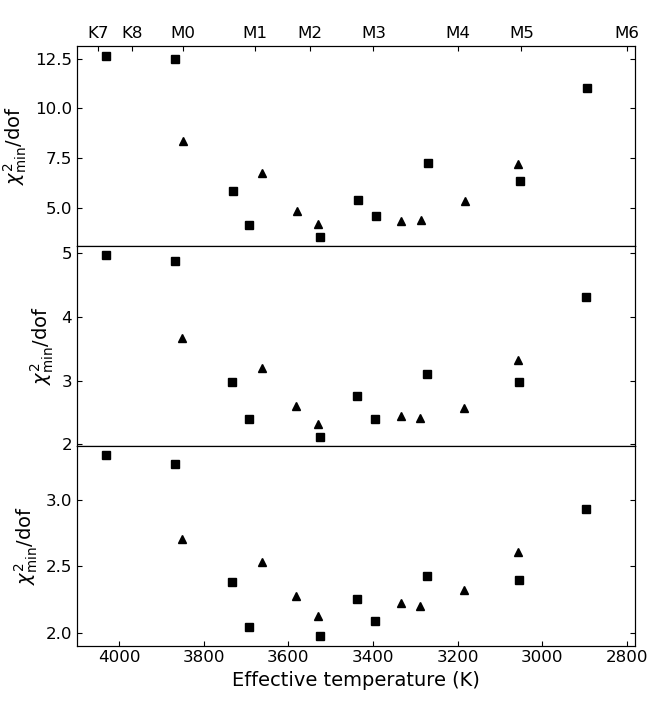}
\caption{\label{fig:spc-class} Results from the spectral classification of the donor star in \targ\ using the optimal subtraction technique with the subsolar templates listed in Table~\ref{tab:templates_rojas_lowFe} (squares) and the supersolar templates from Table~\ref{tab:templates_rojas_highFe} (triangles). The top panel shows the results for the full average spectrum of YY~Dra, while the middle and bottom panels are for the averages at orbital phases $0.9-0.1$ and $0.4-0.6$, respectively. The relation between spectral type and effective temperature is illustrated at the top.}
\end{figure}

For the spectral classification, we applied the optimal subtraction technique outlined in \cite{marsh-94} combining all the YY~Dra spectra, including the 2020 February dataset. This method optimises the subtraction of a scaled and broadened spectral template from the Doppler-corrected average spectrum of the target (see Fig.~\ref{fig:optsub} for an illustration of the technique). We detail here how we proceeded with each of the templates. First, we velocity-shifted all \targ\ and template spectra to a common rest frame. Subsequently, we computed a weighted average of the \targ\ spectra, with weights proportional to the S/N. We then broadened the absorption lines of the template spectrum through convolution with Gray's rotational profile \citep{libro-gray}, adopting a linear limb-darkening coefficient of $0.19$ for the $K$ band, which is a plausible value for an early M-type star \citep{claret95}. The $v_\mathrm{rot} \sin i$ space was explored between $1$ and $200$\,\kms\ in steps  of $1$\,\kms . To account for the fact that the donor star contributes a fraction, $f$, to the total flux in our wavelength range, the broadened versions of the template were multiplied by different $f$ values $> 0$. Finally, all these broadened and scaled versions of the template were subtracted from the \targ\ average spectrum. The optimal values for the broadening and $f$ are those that minimise the $\chi^2$ between the residual of the subtraction and a smoothed version of itself, obtained by convolution with a $\mathrm{FWHM}=30$~\AA\ Gaussian. The minimum $\chi^{2}/\mathrm{dof}$ for each template is listed in Tables~\ref{tab:templates_rojas_lowFe} and~\ref{tab:templates_rojas_highFe}, and the top panel of Fig.~\ref{fig:spc-class} illustrates these results. The robust measurement of $v_\mathrm{rot} \sin i$ will be presented in Sect.~\ref{sec-rotationalbroadening}, where we employ spectral type templates taken with the same instrumental setup as the YY~Dra spectra. 

Considering both grids as a whole, the $\chi^{2}/\mathrm{dof}$ exhibits a plateau between $\simeq 3250$ and $\simeq 3700$~K, corresponding to an spectral type range of M0.5--M3.5 according to the tabulation for main sequence stars by \cite{pecaut-13}. In particular, the best-fit template has an effective temperature of $\simeq$~$3530$~K for both grids, corresponding to an M2~V star. This is in agreement with the M1.8~V classification by \cite{harrison-16}, which comes from a single $K$-band spectrum taken at orbital phase $0.86$. Nevertheless, UV/X-ray irradiation could result in an earlier spectral type in the donor's dayside. To investigate this, we produced average spectra of \targ\ for orbital phases $0.9-0.1$ (night hemisphere, 9 individual spectra) and $0.4-0.6$ (day hemisphere, 10 individual spectra) and repeated the optimal subtraction procedure for both. The results are depicted in the middle and bottom panels of Fig.~\ref{fig:spc-class}. In both cases, the preferred effective temperature range remains very similar, and the best-fit subsolar and supersolar templates align with that from the full average. This suggests that the temperatures of the day and night sides of the donor star must be quite close. We also searched for a narrow component in the Br$\gamma$ emission line profiles coming from a heated hemisphere of the donor star, but it was not detected.

The $f$ values fall in the $\simeq 0.8 - 1.0$ range. The best-matching templates in the supersolar grid (Table~\ref{tab:templates_rojas_highFe}) exhibit deeper absorption
lines, resulting in a lower $f$ measurement  compared to templates of similar effective temperatures in the subsolar grid (Table~\ref{tab:templates_rojas_lowFe}). The latter give $f = 1$ or even higher, which is not physically possible. Two features in our data make clear that $f < 1$: the flickering in the light curves (Fig.~\ref{fig:lightcurves}) and the presence of the Br$\gamma$ emission line in the spectra (Fig.~\ref{fig:optsub}), with an equivalent width of $-33 \pm 7$~\AA\ (mean and standard deviation).\footnote{For comparison, the equivalent width of the Br$\gamma$ emission line is $-14 \pm 6$~\AA\ in the spectroscopy of XY Ari when $f \simeq 0.7 - 0.8$ in the $K$ band \citep{ayoze-2023}.} Moreover, our modelling of the data (presented in Sect.~\ref{sec-modelling}) supports that the donor star contributes $\simeq 80 - 85\%$ to the $K$-band flux, in line with the values obtained with the best-fit templates with supersolar metallicity (Table~\ref{tab:templates_rojas_highFe}). These arguments suggest that the metallicity of the donor star in YY~Dra is probably higher than solar, in contrast with the $\mathrm{[Fe/H]} = -0.20$~dex claimed by \cite{harrison-16}. We attribute this disagreement to the fact that \cite{harrison-16} did not consider the reduction in the depth of the absorption lines produced by veiling. In fact, \cite{harrison-2018} concludes that many of the CV spectra identified as having low metallicity in \cite{harrison-16} are probably affected by non-negligible veiling. In this context, \cite{harrison-2018} obtains $\mathrm{[Fe/H]} = 0.0$~dex for SS Cyg and RU Peg when accounting for the plausible veiling, while they were reported to have $\mathrm{[Fe/H]} = -0.3$~dex in \cite{harrison-16}.

\subsection{Projected rotational velocity of the donor star}\label{sec-rotationalbroadening}

For the analysis in this section, we once again employed the optimal subtraction technique. This time we used the three spectral templates of M-type dwarfs obtained during our observational campaign: Gl~176, Gl~797~B and LHS~3558. The effective temperatures of these templates (see Table~\ref{tab:vsini}) are within the constraints found in Sect.~\ref{sec-spectralclassification} for the donor star of \targ . Since these templates are subject to the same instrumental broadening as the target spectra, using the optimal subtraction method with them allow us to achieve a robust measurement of $v_\mathrm{rot} \sin  i$. Prior to this, we simulated in our template spectra the impact of the radial velocity drift in the absorption lines during the length of the spectroscopic observations: for each template, we generated as many copies as the number of target spectra. Then, we applied to these copies the smearing suffered by the individual YY~Dra spectra through convolution with a rectangular profile of the corresponding width, calculated from the effective length of the exposures and the orbital parameters. Finally, these smeared copies were averaged using identical weights as for the target spectra. The smearing correction is negligible: not applying it results in the exact same values and errors for $v_\mathrm{rot} \sin  i$ and $f$.

As part of the analysis, we examined the impact on the value of $v_\mathrm{rot} \sin  i$ of the FWHM of the Gaussian employed to smooth the residual resulting from the optimal subtraction. We find that varying the FWHM between 1 and 50~\AA\ implies changes in the value of $v_\mathrm{rot} \sin  i$ below $5$~\kms , displaying a weak parabolic relationship with a minimum at $\simeq 30$~\AA. \cite{ayoze-2023} conducted tests on data taken with the same instrumental configuration that demonstrate that the minimum in the FWHM-broadening curves represents the actual $v_\mathrm{rot} \sin  i$. Consequently, we select the $v_\mathrm{rot} \sin  i$ values derived using a 30-\AA \, FWHM smoothing Gaussian.

To assess the uncertainties in $v_\mathrm{rot} \sin  i$ and $f$, we employed the Monte Carlo procedure described in \cite{steeghs-2007} and \cite{torres-2020}. We generated 10000 bootstrapped copies of the average spectrum of \targ , repeating the optimal subtraction technique for each of them. This process yielded probability distributions for $v_\mathrm{rot} \sin  i$ and $f$, both of which were effectively fitted by Gaussians. Therefore, in both cases, we adopted the mean and standard deviation as the representative value and the $1 \, \sigma$ uncertainty, respectively.

Table~\ref{tab:vsini} presents the $v_\mathrm{rot} \sin  i$ and $f$ values obtained using the three spectral templates. Their metallicities are in the range  $-0.23 \leq \mathrm{[Fe/H]} \leq 0.15$~dex, having a negligible effect on $v_\mathrm{rot} \sin  i$. The mean $v_\mathrm{rot} \sin  i$ is $103$~\kms . For subsequent sections, we adopt this value while retaining the uncertainty of the individual measurements ($103 \pm 2$~\kms). This decision partially accounts for potential systematics introduced by the choice of the linear limb-darkening coefficient in Gray's rotational profile \citep{libro-gray}. In this respect, we have used a limb-darkening coefficient which is suitable for the continuum \citep{claret95}, but absorption lines in late-type stars have lower limb-darkening than the continuum \citep{collins-95}. Adopting the extreme case of zero limb-darkening changes the $v_\mathrm{rot} \sin  i$ value by $-1$~\kms .

\begin{table}
\caption[]{$v_\mathrm{rot} \sin  i$ and $f$ values for YY~Dra. }
\centering
\begin{tabular}{lcccccc}
\hline\noalign{\smallskip}
Template & Sp. & Effective & $v_\mathrm{rot} \sin  i$ & $f$  \\
\tabhead{} & Type & temperature & (\kms) & \\
  &  & (K) & & \\ 
\hline\noalign{\smallskip}
Gl 176 & M2 V & $3581 \pm 20$  & $102(2)$ & $0.79(2)$ \\
Gl 797 B & M2 V & $3569 \pm 20$  & $104(2)$ & $0.97(2)$ \\
LHS 3558 & M3 V & $3437 \pm 16 $  & $104(2)$ & $0.89(2)$  \\
\hline\noalign{\smallskip}
\end{tabular}
\label{tab:vsini}
\tablefoot{We derived $v_\mathrm{rot} \sin  i$ and $f$ through optimal subtraction of template spectra taken with the same instrumental setup as used for the target. Numbers in brackets indicate the uncertainty in the last digit. The effective temperatures of the templates are from \cite{rojas-ayala-2012}.}
\end{table}

The modulation of $v_\mathrm{rot} \sin  i$ with orbital phase is well-documented (see e.g. \citealt{tariq-14}). Our spectra of \targ\ cover the full orbital cycle with a consistently uniform sampling. As a result, the $v_\mathrm{rot} \sin  i$ measurements provided above are expected to closely represent the mean value throughout the entire orbit. We verified this by checking that a consistent result (within $1 \, \sigma$ agreement) is obtained when averaging the $v_\mathrm{rot} \sin  i$ measurements at orbital phases $0.9-0.1$, $0.15-0.35$, $0.4-0.6$, and $0.65-0.85$.

\subsection{Modelling of the light and radial velocity curves}\label{sec-modelling}
To determine the fundamental parameters of YY~Dra, we conducted simultaneous modelling of the two $K_{\mathrm{s}}$-band light curves (Sect.~\ref{subsec-photometry}) and the 2021~May~+~2022~May radial velocities of the donor star (Sect.~\ref{sec-k2}). We selected only these radial velocities since both of them exhibit a consistent behaviour (see the text in Sect.~\ref{sec-k2}). All the curves were phase-folded using the ephemeris provided by \cite{katherine-22}, and the light curves were averaged into $40$ orbital phase bins, as shown in the bottom panel of Fig.~\ref{fig:lightcurves}. We used \textsc{XRBCURVE}, a code described in \cite{tariq-2000,tariq-2003,tariq-2017}. The model considers a binary system composed of a primary star (assumed to be a point source) surrounded by an accretion disc and a co-rotating secondary star. The Roche lobe-filling factor of the latter can be specified, and we set it to $1$. The accretion disc is circular and has a height $H$ at the outer edge, and we adopted the circularization radius \citep{warner-libro} as the disc outer radius. We assumed that the accretion disc and the WD provide constant flux through the whole orbit, as no evidence of a hotspot or similar feature was found in the light curves (Sect.~\ref{subsec-photometry}). 

In the absence of irradiation, the flux of the donor star depends only on its unperturbed effective temperature ($T_2$) and its limb- and gravity-darkening. We calculated its intensity distribution using \textsc{NEXTGEN} model-atmosphere fluxes \citep{hauschildt-99} and the tabulations by \cite{claret-2011} for the limb- and  gravity-darkening coefficients. To account for the irradiation of the donor star’s surface, we followed \cite{tariq-2003} and \cite{tariq-2017}. Firstly, we computed the increase in the effective temperature for each surface element due to the UV/X-ray irradiation, adopting an UV/X-ray albedo of 0.5 for the donor star. Using albedo values of $0.1$ and $1$ produces consistent results within $1 \, \sigma$ for all the parameters and changes $\lesssim 1$ per~cent in the derived masses.\footnote{For zero albedo, the
masses of the WD and the donor star increase by $\simeq 5 \%$ and $\simeq 4 \%$, respectively. However, this has no impact on the discussion presented in Sect.~\ref{sec-discuss}.} Subsequently, we integrated the entire visible part of the donor star at each orbital phase to determine the total flux. For the radial velocity curve, the approach was similar: we used an equivalent width-effective temperature relation derived from \textsc{PHOENIX} model spectra \citep{husser-2013} to establish the equivalent width of the absorption lines for each surface element based on its effective temperature. To account for the fact that absorption lines can disappear in surface elements that are subject to high irradiation levels, we introduced the limiting factor $F_{\mathrm{AV}}$. This represents the maximum fraction of total-to-unperturbed flux that a surface element can have while still maintaining the absorption features. For instance, if $F_{\mathrm{AV}} = 1.50$, the equivalent width of the lines is set to zero in the surface elements where the incident flux exceeds $50$ per~cent of the unperturbed flux. Finally, we integrated the entire visible part of the donor star at each orbital phase to determine the corresponding line-of-sight radial velocity.

\begin{figure*}
\sidecaption
\centering \includegraphics[width=12cm]{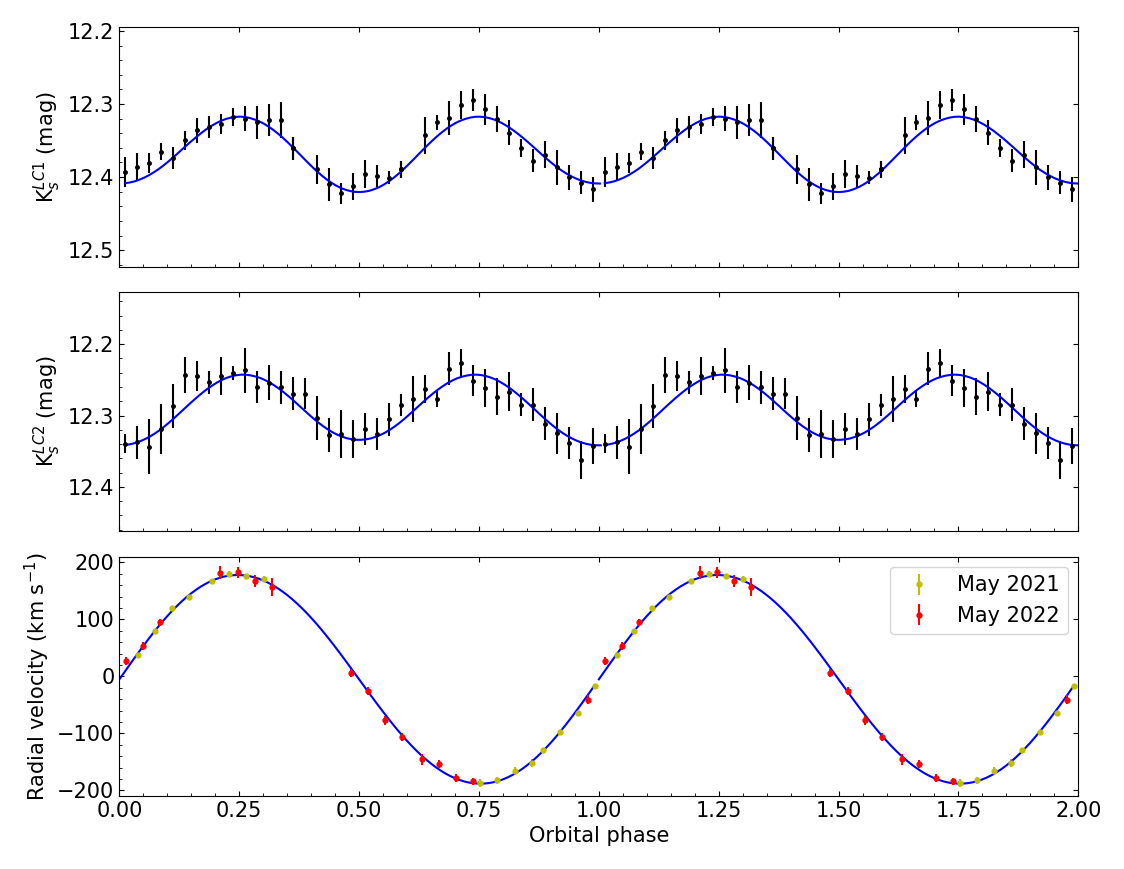}
\caption{\label{fig:fits_modelling} Result of the simultaneous modelling of the light and radial velocity curves of YY Dra. From top to bottom: phase-binned $K_{\mathrm{s}}$-band light curves from 2021 March 31 and May 5, followed by the radial velocity curve of the donor star (excluding the 2020 February data). The blue solid lines represent the best fit. The orbital cycle is shown twice for the sake of clarity. The fit of the first light curve (top panel) displays the absolute minimum at orbital phase $0.5$, while it is at phase $0$ in the fit of the second curve (middle panel) due to changes in irradiation ($F_{X,0}^{\mathrm{LC2}} > F_{X,0}^{LC1}$, Table~\ref{tab:parameters_lcm_mcmc}).}
\end{figure*}

\begin{figure*}
\centering \includegraphics[height=18.5cm]{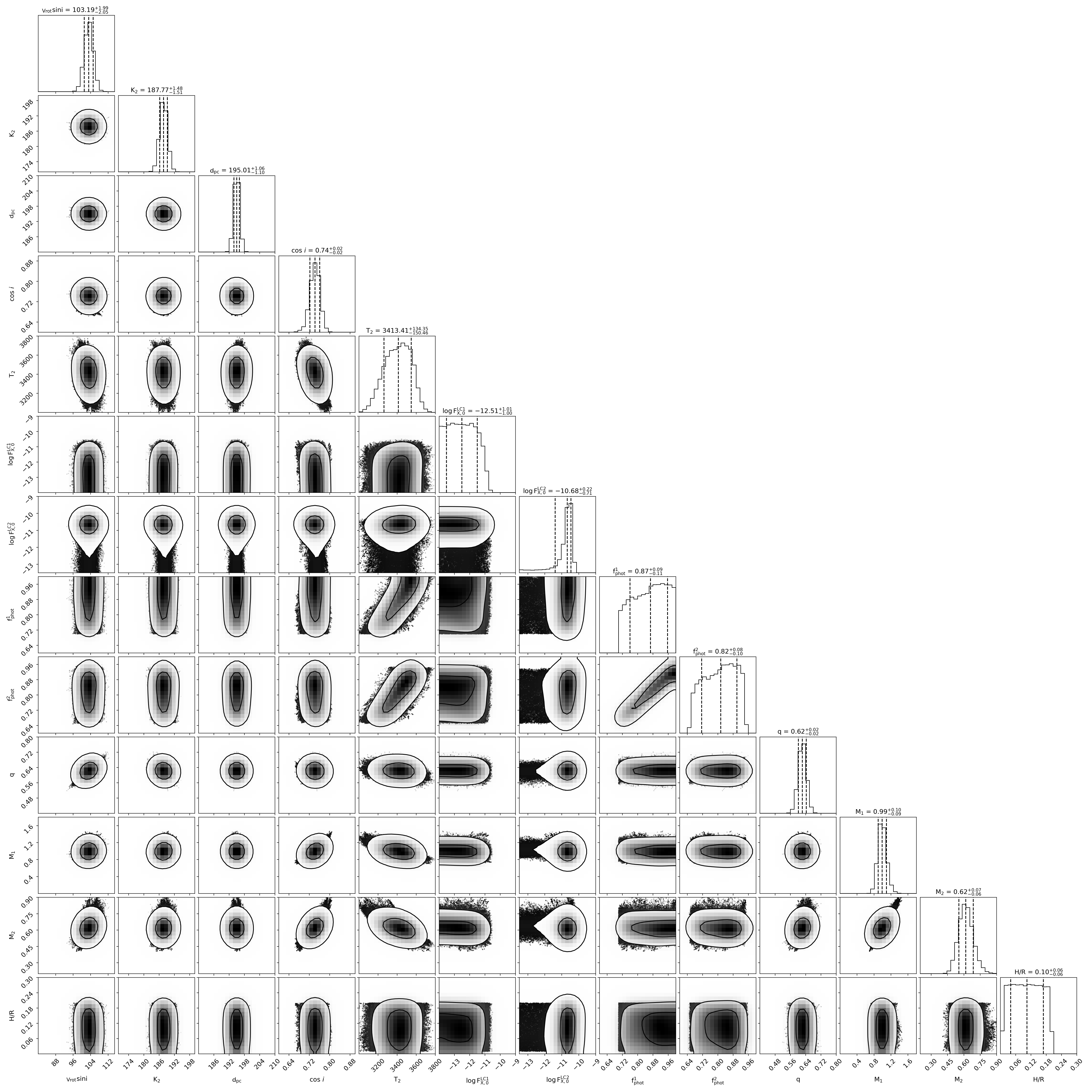}
\caption{\label{fig:cornerplot} Correlation diagrams illustrating the probability distributions resulting from the simultaneous modelling of the light and radial velocity curves of YY~Dra using \textsc{XRBCURVE}. Only the most relevant parameters are shown. The $68$, $95$ and $99.7$ per cent confidence contours are indicated in the 2D plots. In the right panels, the projected 1D parameter distributions are presented, with dashed lines indicating the mean and $68$ per cent confidence level intervals.}
\end{figure*}

The free parameters of the model include $\gamma$, $K_2$, $\cos i$, $v_\mathrm{rot} \sin  i$, $T_2$, the distance to the source in parsecs ($d_{\mathrm{pc}}$) and the height-to-radius ratio of the accretion disc ($H/R$). Additionally, we parameterise the unabsorbed UV/X-ray heating flux of the WD for both the first ($F_{X,0}^{\mathrm{LC1}}$) and second ($F_{X,0}^{\mathrm{LC2}}$) light curves, where the latter also represents the UV/X-ray flux for the radial velocity curve due to simultaneity. To properly model irradiation, $F_{\mathrm{AV}}$ (described above) is another free parameter. The flux contribution to the $K_{\mathrm{s}}$ band from the WD and the accretion disc introduce two additional parameters, represented in terms of the donor star's fractional contribution to the $K_{\mathrm{s}}$-band flux, denoted as $f^{1}_{\mathrm{phot}}$ and $f^{2}_{\mathrm{phot}}$ for the first and second light curves, respectively. Finally, the set of free parameters is completed by phase shifts of the synthetic curves with respect to the observed curves. These are $\delta \phi_{\mathrm{LC1}}$ and $\delta \phi_{\mathrm{LC2}}$ for the first and second light curves, respectively, and $\delta \phi_{\mathrm{RVC}}$ for the radial velocity curve. They account for uncertainties in the ephemeris and for potential differences between the spectroscopic and the photometric $T_0$.

To simultaneously fit the two light curves and the radial velocity curve, we used a Markov chain Monte Carlo (MCMC) method convolved with a differential evolution fitting algorithm within a Bayesian framework (see \citealt{tariq-2017} and references within). We utilised flat prior probability distributions for $\gamma$, $K_2$, $\mathrm{cos} \, i$, $T_2$, $H/R$, $F_{X,0}^{\mathrm{LC1}}$, $F_{X,0}^{\mathrm{LC2}}$, $F_{\mathrm{AV}}$, $f^{1}_{\mathrm{phot}}$, $f^{2}_{\mathrm{phot}}$, $\delta \phi_{\mathrm{LC1}}$, $\delta \phi_{\mathrm{LC2}}$ and $\delta \phi_{\mathrm{RVC}}$. For $T_2$, we employed the $3250 - 3700$~K limits, according to the results from Sect.~\ref{sec-spectralclassification}. We imposed Gaussian priors in $v_\mathrm{rot} \sin  i$ and $d_{\mathrm{pc}}$ based on our result from Sect.~\ref{sec-rotationalbroadening} and the distance from \cite{bailer-jones-21}, respectively. We employed 60 walkers to explore the parameter space and 15000 steps per chain, discarding the first 1000 as burn-in. The best simultaneous fit to the light and radial velocity curves is displayed in Fig.~\ref{fig:fits_modelling}, while Fig.~\ref{fig:cornerplot} presents the correlation plot. At each iteration, $q$ was calculated using $v_\mathrm{rot} \sin  i$ and $K_2$ following equation 2 in \cite{ayoze-2023}, while the masses were derived from $q$, $K_2$, $i$ and $P$ applying equations 5 and 6 in the same reference. The $\chi^2$ value for the simultaneous fit is $53.7$ (dof = 97). Table~\ref{tab:parameters_lcm_mcmc} provides the resulting model parameters and the $1 \, \sigma$ statistical uncertainties.

\begin{table}
\caption[]{Best-fit parameters of the simultaneous modelling of the light and radial velocity curves using \textsc{XRBCURVE}.}
\label{tab:parameters_lcm_mcmc}
\centering
\begin{tabular}{l l c }
\hline\noalign{\smallskip}
Parameter & Prior & Value \\
\hline\noalign{\smallskip}
$\gamma$ ($\mathrm{km}\,\mathrm{s}^{-1}$) & [-30.0, 10.0] & $-5 \pm 1$\\
$K_2$ ($\mathrm{km}\,\mathrm{s}^{-1}$) & [160, 250] & $188^{+1}_{-2}$\\
\noalign{\smallskip}
$\mathrm{cos} \, i$ & [0.5, 0.99] & $0.74 \pm 0.02$\\
\noalign{\smallskip}
$v_\mathrm{rot} \sin  i$ ($\mathrm{km}\,\mathrm{s}^{-1}$) & $103 \pm 2$ & $103 \pm 2$\\
\noalign{\smallskip}
$d_{\mathrm{pc}}$ & $195 \pm 1$ & $195 \pm 1$\\
\noalign{\smallskip}
$T_2 (\mathrm{K})$ & [3250, 3700] & $3413^{+134}_{-150}$\\
\noalign{\smallskip}
$\mathrm{log} \, F_{X,0}^{LC1} \mathrm{(\mathrm{erg}\,\mathrm{s}^{-1} \,  \mathrm{cm}^{-2})}$ & [-14.0, -9.0] & $-12.5 \pm 1.0$\\
\noalign{\smallskip}
$\mathrm{log} \, F_{X,0}^{LC2} \mathrm{(\mathrm{erg}\,\mathrm{s}^{-1} \,  \mathrm{cm}^{-2})}$ & [-14.0, -9.0] &  $-10.7^{+0.2}_{-0.7}$\\
\noalign{\smallskip}
$f^{1}_{\mathrm{phot}}$  & [0.3, 1.0] &  $0.87^{+0.09}_{-0.11}$\\
\noalign{\smallskip}
$f^{2}_{\mathrm{phot}}$  & [0.3, 1.0] & $0.82^{+0.08}_{-0.10}$\\
\noalign{\smallskip}
$H/R$ & [0.01, 0.2] & $0.10 \pm 0.06$\\
\noalign{\smallskip}
$F_{\mathrm{AV}}$ & [1.0, 5.0] & $3.0 \pm 1.3$ \\
\noalign{\smallskip}
$\delta \phi_{\mathrm{LC1}}$ & [-0.15, 0.15] & $0.012 \pm 0.005$\\
\noalign{\smallskip}
$\delta \phi_{\mathrm{LC2}}$ & [-0.15, 0.15] & $0.014 \pm 0.008$\\
\noalign{\smallskip}
$\delta \phi_{\mathrm{RVC}}$ & [-0.15, 0.15] & $0.000 \pm 0.001$\\
\hline\noalign{\smallskip}
\end{tabular}
\tablefoot{Priors indicated between brackets are flat, while priors expressed as value~$\pm$~uncertainty are Gaussian.}
\end{table}

In our model, the accretion disc shadowing over the donor star surface is determined by the $H/R$ ratio, for which we find a near uniform posterior. To evaluate if shadowing could have any impact on our calculations, we repeated the fit adopting a flat disc ($H = 0$), resulting in changes in the parameters $\lesssim 1$ per~cent. The result from this test alleviates the concerns about potential systematic errors induced by simplifying the geometry of the accretion structures in YY~Dra to a disc. It also addresses the fact that shadowing may be lower due to the vertical extent of the irradiating regions. Moreover, shadowing only plays an important role in systems where irradiation is sufficiently high (see e.g. \citealt{shahbaz-2019}). In this regard, our model predicts changes $<20$~K between the day and night hemispheres of the donor star, which is an indication of low heating. Accordingly, none of the surface elements reaches the best-fit limiting factor $F_{\mathrm{AV}}$ (Table~\ref{tab:parameters_lcm_mcmc}). We also find that there are not systematic errors introduced by our selection of the radial velocities for the modelling: using only the 2021 May velocities yields the same results (change $\lesssim 1\%$) for all the parameters, while including the 2020 February data provides consistent values (within $1 \, \sigma$).

The $K_2$ derived from the modelling is $188^{+1}_{-2}$~\kms . For comparison, \cite{mateo-91} obtained $K_2 = 184 \pm 10$~\kms\ . To test our result, we followed \cite{marsh-88} and fitted a circular orbit to the radial velocities in the orbital phase range $0.8 - 1.2$. We derived $K_2 = 188 \pm 3$~\kms , which matches the value obtained from the modelling. The possible $K_2 > K_{\mathrm{abs}}$ may be a consequence of irradiation being sufficient to increase the brightness of the day side without changing the spectral type. This shifts the centre of light of the absorption line spectrum towards the WD, as proposed to explain the radial velocity curve of SS Cyg when irradiation is low \citep{robinson-86,bitner-2007}. The mass ratio is $q=0.62 \pm 0.02$, which significantly differs from the previously reported values of $q=0.48 \pm 0.07$ and $q=0.45 \pm 0.05$ presented in \cite{mateo-91} and \cite{haswell-97}, respectively. These were derived from the ratio of radial velocity semi-amplitude of emission and absorption lines. In constrast, our value agrees with $q=0.61 \pm 0.09$, derived by \cite{haswell-97} using the $v_\mathrm{rot} \sin  i = 110 \pm 10$~\kms\ and $K_{\mathrm{Na \textsc{i}}} = 202 \pm 3$~\kms\ measurements by \cite{friend-90}. The resulting $\cos \, i = 0.74 \pm 0.02$ yields $i={42^{\circ}}^{+2^{\circ}}_{-1^{\circ}}$. This is consistent with previous estimates in the literature: $i = 42^{\circ} \pm 5^{\circ}$ \citep{mateo-91}, $i = 45^{\circ} \pm 4^{\circ}$ \citep{haswell-97} and $i = 41^{\circ} \pm 3^{\circ}$ \citep{joshi-2012}. The derived binary masses are:
\par
~\par
$M_{1} = 0.99^{+0.10}_{-0.09} \, \mathrm{M}_{\odot}$~, \, \, \, \, \, \, \, \, \, \, \, \, \,  $M_2 = 0.62^{+0.07}_{-0.06} \, \mathrm{M}_{\odot}$~.

\comment{We must note that our model did not include an accretion disc for several reasons. First, there are not eclipses of any type. Second, this approach reduces the number of free parameters. Third, some of the disc properties change over time and only could be estimated through rough assumptions (radius, height, profile of temperature, etc). Finally, the accretion structures of an intermediate polar are not properly modelled by a disc. However, since we are not including a disc, our model can not reproduce the potential X-ray/UV shadowing in the region of the donor star near the inner Lagrangian point (see e.g. \citealt{phillips-99}). Fortunately, this is unlikely to have a significant impact in YY~Dra because the irradiating flux is not enough to produce a detectable difference in the effective temperature of the day and night sides of the donor star (Sect.~\ref{sec-spectralclassification}). In fact, we decided to check if shadowing could affect the parameters by repeating the modelling including an accretion disc with opening angle $\alpha$ as a free parameter. This test produced the same values for all the parameters presented in Table~\ref{tab:parameters_lcm_mcmc} (within $1 \, \sigma$).}

\section{Discussion}\label{sec-discuss}

\subsection{Derived properties of the stellar components}

The spectrum of the donor star in \targ\ is consistent with a spectral type M0.5--M3.5~V, most likely $\simeq$M2~V (Sect.~\ref{sec-spectralclassification}). This finding is supported by the previous classification of M1.8~V using the same spectroscopic band by \cite{harrison-16}. The dynamical mass is $M_{2}= 0.62^{+0.07}_{-0.06} \, \mathrm{M}_{\odot}$, and the Roche lobe volume radius can be calculated as $R_2 = \frac{P}{2 \pi} \frac{v_\mathrm{rot} \sin  i} {\mathrm{sin} \, i} = 0.50 \pm 0.02 \, R_{\odot}$. The surface gravity $\log \, g = 4.83 \pm 0.06$~dex derived from these quantities falls within the expected range for a dwarf star. According to \cite{pecaut-13}\footnote{\url{https://www.pas.rochester.edu/~emamajek/EEM_dwarf_UBVIJHK_colors_Teff.txt}}, an M2~V star has a mass of $\simeq 0.44 \, \mathrm{M}_{\odot}$. This suggests that the donor star in YY~Dra is more massive than expected for its spectral type, as observed in the dwarf nova IP Peg ($P \simeq 3.80$~h, \citealt{copperwheat-2010}) and the nova-like CVs HS~0220+0603 ($P \simeq 3.58$~h, \citealt{pablo-2015}) and KR Aur ($P \simeq 3.91$~h, \citealt{pablo-2020}). Higher masses have also been found in non-interacting dwarfs: there are robustly-classified M1-M3~V stars with accurately determined masses of $\simeq 0.6 - 0.7 \, \mathrm{M}_{\odot}$ (see e.g. tables~2 and~5 in \citealt{guilletorres-2010}).

Donor stars in CVs are supposed to be slightly oversized with respect to isolated main sequence stars of the same mass (see e.g. \citealt{patterson-2005,knigge-2006,knigge-2011}). This is due to the thermal timescale, $\tau_{\mathrm{Th}}$, being similar to the mass-loss timescale, $\tau_{\mathrm{ML}}$. Using our dynamical mass determination, the empirical mass–radius relation for donor stars in CVs derived by \cite{knigge-2011} indicates a radius of $0.64 \pm 0.08 \, R_{\odot}$, which only agrees within $2 \, \sigma$ with our $R_2$ value. Adopting that the accretion rate for YY~Dra ($\simeq 3.5 \times 10^{15}$~$\mathrm{g} \, \mathrm{s}^{-1}$, \citealt{suleimanov-19}) is equal to the donor mass-transfer rate, $\dot{M_2}$, we calculate $\tau_{\mathrm{ML}} \simeq M_2 / \dot{M_2} \simeq 1.3 \times 10^{10}$~yr and $\tau_{\mathrm{Th}} \simeq G \, M_2^{2} \, / \, L_2 \, R_2 \simeq 6.7 \times 10^{8}$~yr, where $L_2 = 4 \pi \sigma R_2^2 \, T_2^4$. These numbers, together with the value of $R_2$, suggest that the donor star in YY~Dra is able to maintain thermal equilibrium in spite of the mass transfer. Similar conclusions have been drawn for IP Peg \citep{copperwheat-2010} and HS~0220+0603 \citep{pablo-2015} based on comparable reasoning.

\comment{This suggests that the spectral type of the donor star in YY~Dra is probably later than expected based on the orbital period, a common feature observed in many CVs (\citealt{Thorstensen-2004,harrison-2018}; see figure~30 in \citealt{harrison-16}).}

For the WD, we derive a dynamical mass of $M_{1} = 0.99^{+0.10}_{-0.09} \, \mathrm{M}_{\odot}$. This is significantly higher than the mean value of $\simeq 0.6 \, \mathrm{M}_{\odot}$ obtained spectroscopically from isolated WDs in the Sloan Digital Sky Survey \citep{kepler-2016,kepler-2019}. However, accurate mass measurements of WDs in CVs suggest they are typically more massive than their isolated counterparts, with an average value of $0.81^{+0.16}_{-0.20} \, \mathrm{M}_{\odot}$ \citep{zorotovic-2011,pala-2022}. Our dynamical mass falls within 1 sigma with the latter.

\subsection{Comparison of the dynamical WD mass with estimates from X-ray spectral studies}\label{subsec-m1x}

Numerous X-ray studies have provided estimates of the mass of WDs in IPs (hereafter $M_1^\mathrm{X}$). The predominant approach for this is modelling the X-ray spectral continuum, as outlined in studies like \cite{shaw-2020}, which relies on certain assumptions. Firstly, it assumes that the temperature of the plasma in the region where the accreted material impacts the WD surface is a function of the WD mass and radius \citep{aizu-73}. Secondly, the main cooling mechanism is presumed to be the emission of X-rays, specifically thermal bremsstrahlung \citep{wu-94,cropper-98}, although some authors have proposed comptonization as a possibility (e.g. \citealt{maiolino-21}). Typically, this method is considered reliable when using hard ($>20$~keV) X-ray spectra, as for masses $>0.6$\,M$_{\odot}$, the maximum temperature exceeds $20$~keV \citep{suleimanov-2005}. Moreover, the soft spectrum of many IPs is significantly affected by incompletely understood absorption and reflection components \citep{suleimanov-19}. Fortunately, all the $M_1^\mathrm{X}$ values from X-ray spectral modelling that we have found in the literature for \targ\ were derived from hard X-ray data. \cite{suleimanov-2005} fitted a $3-100$~keV \textit{RXTE}/PCA+HEXTE spectrum using a multi-temperature bremsstrahlung model with the approximation of accreted matter falling from infinity, obtaining $M_1^\mathrm{X} = 0.75 \pm 0.05 \, \mathrm{M}_{\odot}$. However, this result only agrees within the $2 \sigma$ level with our dynamical measurement. \cite{brunchsweiger-2009} employed the same model on $15-195$~keV \textit{Swift}/BAT data, obtaining a lower estimate, $M_1^\mathrm{X} = 0.50 \pm 0.11 \, \mathrm{M}_{\odot}$. They suggested that their value might be an underestimate due to the inappropriateness of the accreted matter falling from infinity assumption for systems with short spin periods. \cite{yuasa-2010} used a similar model with $3-50$~keV \textit{Suzaku} data and found $M_1^\mathrm{X} = 0.67^{+0.30}_{-0.14} \, \mathrm{M}_{\odot}$, in agreement with our dynamical mass, but with high uncertainty. Using a comparable approach, \cite{xu-2019} modelled $0.3-50$~keV \textit{Suzaku} spectroscopy and derived $M_1^\mathrm{X} = 0.82 \pm 0.04 \, \mathrm{M}_{\odot}$, which agrees within $2 \sigma$ with our measurement. On the other hand, \cite{suleimanov-19} used a more complex model considering a finite fall height for the accreted matter and fitted a $15-195$~keV \textit{Swift}/BAT spectrum, obtaining $M_1^\mathrm{X} = 0.76 \pm 0.08 \, \mathrm{M}_{\odot}$. This falls below our dynamical WD mass but remains within a $2 \sigma$ agreement. 

Other X-ray studies try to infer the WD mass from emission lines rather than fitting the spectral continuum. \cite{xu-2019} employed synthetic X-ray spectra to establish a relation between the ratio of the $6.7$~and $7.0$~keV iron emission lines and the WD mass. By measuring that ratio in \textit{Suzaku} spectroscopy and applying their relation, they found $M_1^\mathrm{X} = 0.80 \pm 0.14 \, \mathrm{M}_{\odot}$. This estimate shows a $1 \sigma$ agreement with our dynamical mass.

\begin{figure}
\centering \includegraphics[height=10cm]{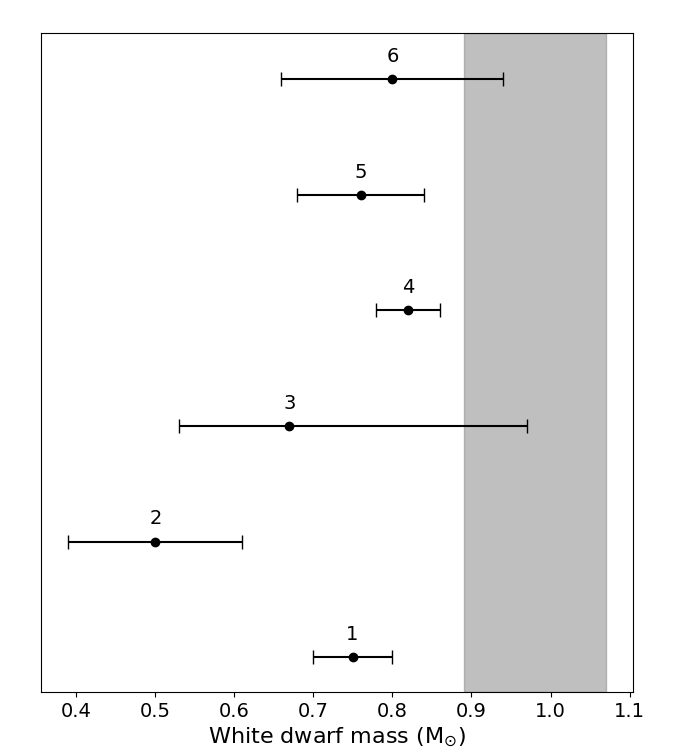}
\caption{\label{fig:masses} Visual comparison of our dynamical WD mass for \targ\ (shaded column, $1 \, \sigma$) with previous estimates derived from spectral modelling of: \textit{RXTE} hard X-ray data (labelled in the figure as $1$; \citealt{suleimanov-2005}), \textit{Swift} hard X-ray data (labelled as $2$ and $5$; \citealt{brunchsweiger-2009} and \citealt{suleimanov-19}) and \textit{Suzaku} hard X-ray data (labelled as $3$ and $4$; \citealt{yuasa-2010} and \citealt{xu-2019}). We also include one estimate based on the ratio of the iron emission lines in \textit{Suzaku} spectroscopy (labelled as $6$; \citealt{xu-2019}).}
\end{figure}

In conclusion, the majority of $M_1^\mathrm{X}$ values derived from modelling of the hard X-ray spectral continuum are likely underestimates. The $M_1^\mathrm{X}$ obtained from the ratio of the $6.7$~and $7.0$~keV iron emission lines shows a $1 \sigma$ agreement with the dynamical value, suggesting a potential correlation between that ratio and the WD mass. However, this method has not been tested against a significant number of dynamical masses. The only previous case is XY Ari, where the high uncertainty of the $M_1^\mathrm{X}$ value prevented any conclusion. Figure~\ref{fig:masses} provides a visual summary of the comparison conducted in this section. Our findings are consistent with those in \cite{ayoze-2021} and \cite{ayoze-2023}, where a significant number of $M_1^\mathrm{X}$ values derived from modelling the hard X-ray spectral continuum of GK~Per and XY~Ari are found to be underestimates, even with refined models that consider the finite fall height of the accreted material onto the WD.

\section{Conclusions}\label{sec-conclusions}
We conducted a dynamical study of the IP \targ\ using partially simultaneous near-infrared spectroscopy and photometry taken between $2020$ and $2022$. We used the spectra to constrain the spectral type of the donor star to M0.5--M3-5~V, derive its radial velocity curve and measure the rotational broadening $v_\mathrm{rot} \sin  i = 103 \pm 2$~\kms. From the photometry, we derived two light curves, which we modelled together with the radial velocity curve. This analysis yielded the fundamental parameters of the system, including the radial velocity semi-amplitude of the centre of mass of the donor star $K_2 = 188^{+1}_{-2}$~\kms\ , the mass ratio $q = 0.62 \pm 0.02$ and the orbital inclination $i=42^{+2^{\circ}}_{-1^{\circ}}$. These results are consistent with previous estimates from the literature. The resulting dynamical masses for the WD and the donor star are $M_{1} = 0.99^{+0.10}_{-0.09} \, \mathrm{M}_{\odot}$ and $M_2 = 0.62^{+0.07}_{-0.06} \, \mathrm{M}_{\odot}$, respectively. Compared to a typical main sequence star of the same spectral type, the donor star has a slightly larger mass, as observed in other CVs. Likewise, the WD mass aligns with the average value in CVs.

When comparing our dynamical WD mass with estimates from modelling of the hard X-ray spectral continuum, we find significant disagreements, in line with previous dynamical studies exploring this issue in other IPs. Additionally, we compared our value with one estimate derived from the ratio of X-ray iron emission lines, revealing good agreement. The latter may be an accurate technique; however, it warrants further testing to confirm it as a reliable method. This dynamical study provides another valuable piece of evidence to assess the accuracy of current X-ray methods in deriving WD masses for IPs.


\begin{acknowledgements}
We thank the anonymous referee for providing constructive suggestions. Based on observations made with the Gran Telescopio Canarias (GTC), installed at the Spanish Observatorio del Roque de los Muchachos of the Instituto de Astrofísica de Canarias, on the island of La Palma. This article is also based on observations made with the Nordic Optical Telescope, owned in collaboration by the University of Turku and Aarhus University, and operated jointly by Aarhus University, the University of Turku and the University of Oslo, representing Denmark, Finland and Norway, the University of Iceland and Stockholm University at the Observatorio del Roque de los Muchachos, La Palma, Spain, of the Instituto de Astrofisica de Canarias. We express our gratitude to the GTC staff, especially Antonio L. Cabrera Lavers, for their assistance in implementing the spectroscopy presented in this paper. Special thanks to Alina Streblyanska for sharing her expertise on EMIR with us. We also extend our thanks to Nicol\'as Cardiel L\'opez for clarifying certain aspects of \textsc{pyemir}. The use of the \textsc{molly} package developed by Tom Marsh is acknowledged. This work was supported by the Agencia Estatal de Investigaci\'on del Ministerio de Ciencia e Innovaci\'on (MCIN/AEI) and the European Regional Development Fund (ERDF)  under grant PID2021-124879NB-I00. MAPT acknowledges to have received support via a Ram\'on y Cajal Fellowship (RYC-2015-17854). PR-G acknowledges support from the Consejer\'ia de Econom\'ia, Conocimiento y Empleo del Gobierno de Canarias and the European Regional Development Fund (ERDF) under grant with reference ProID2021010132. P.G.J.~has received funding from the European Research Council (ERC) under the European Union’s Horizon 2020 research and innovation programme (Grant agreement No.~101095973). 
\end{acknowledgements}



\bibliographystyle{aa}
\bibliography{bibliography}

\begin{thebibliography}{71}
\expandafter\ifx\csname natexlab\endcsname\relax\def\natexlab#1{#1}\fi

\bibitem[{{Aizu}(1973)}]{aizu-73}
{Aizu}, K. 1973, Progress of Theoretical Physics, 49, 1184

\bibitem[{{{\'A}lvarez-Hern{\'a}ndez} {et~al.}(2021){{\'A}lvarez-Hern{\'a}ndez}, {Torres}, {Rodr{\'\i}guez-Gil}, {Shahbaz}, {Anupama}, {Gazeas}, {Pavana}, {Raj}, {Hakala}, {Stone}, {Gomez}, {Jonker}, {Ren}, {Cannizzaro}, {Pastor-Marazuela}, {Goff}, {Corral-Santana}, \& {Sabo}}]{ayoze-2021}
{{\'A}lvarez-Hern{\'a}ndez}, A., {Torres}, M.~A.~P., {Rodr{\'\i}guez-Gil}, P., {et~al.} 2021, \mnras, 507, 5805

\bibitem[{{{\'A}lvarez-Hern{\'a}ndez} {et~al.}(2023){{\'A}lvarez-Hern{\'a}ndez}, {Torres}, {Rodr{\'\i}guez-Gil}, {Shahbaz}, {S{\'a}nchez-Sierras}, {Acosta-Pulido}, {Jonker}, {Gazeas}, {Hakala}, \& {Corral-Santana}}]{ayoze-2023}
{{\'A}lvarez-Hern{\'a}ndez}, A., {Torres}, M.~A.~P., {Rodr{\'\i}guez-Gil}, P., {et~al.} 2023, \mnras, 524, 3314

\bibitem[{{Andronov} \& {Mishevskiy}(2018)}]{andronov-2018}
{Andronov}, I.~L. \& {Mishevskiy}, N.~N. 2018, Research Notes of the American Astronomical Society, 2, 197

\bibitem[{{Bailer-Jones} {et~al.}(2021){Bailer-Jones}, {Rybizki}, {Fouesneau}, {Demleitner}, \& {Andrae}}]{bailer-jones-21}
{Bailer-Jones}, C.~A.~L., {Rybizki}, J., {Fouesneau}, M., {Demleitner}, M., \& {Andrae}, R. 2021, \aj, 161, 147

\bibitem[{{Bitner} {et~al.}(2007){Bitner}, {Robinson}, \& {Behr}}]{bitner-2007}
{Bitner}, M.~A., {Robinson}, E.~L., \& {Behr}, B.~B. 2007, \apj, 662, 564

\bibitem[{{Brunschweiger} {et~al.}(2009){Brunschweiger}, {Greiner}, {Ajello}, \& {Osborne}}]{brunchsweiger-2009}
{Brunschweiger}, J., {Greiner}, J., {Ajello}, M., \& {Osborne}, J. 2009, \aap, 496, 121

\bibitem[{{Cardiel} {et~al.}(2019){Cardiel}, {Pascual}, {Gallego}, {Cabello}, {Garz{\'o}n}, {Balcells}, {Castro-Rodr{\'\i}guez}, {Dom{\'\i}nguez-Palmero}, {Hammersley}, {Laporte}, {Patrick}, {Pell{\'o}}, {Prieto}, \& {Streblyanska}}]{cardiel-19}
{Cardiel}, N., {Pascual}, S., {Gallego}, J., {et~al.} 2019, in Astronomical Society of the Pacific Conference Series, Vol. 523, Astronomical Data Analysis Software and Systems XXVII, ed. P.~J. {Teuben}, M.~W. {Pound}, B.~A. {Thomas}, \& E.~M. {Warner}, 317

\bibitem[{{Chanmugam} \& {Wagner}(1977)}]{chanmugam-77}
{Chanmugam}, G. \& {Wagner}, R.~L. 1977, \apjl, 213, L13

\bibitem[{{Claret} \& {Bloemen}(2011)}]{claret-2011}
{Claret}, A. \& {Bloemen}, S. 2011, \aap, 529, A75

\bibitem[{{Claret} {et~al.}(1995){Claret}, {Diaz-Cordoves}, \& {Gimenez}}]{claret95}
{Claret}, A., {Diaz-Cordoves}, J., \& {Gimenez}, A. 1995, \aaps, 114, 247

\bibitem[{{Collins} \& {Truax}(1995)}]{collins-95}
{Collins}, George~W., I. \& {Truax}, R.~J. 1995, \apj, 439, 860

\bibitem[{{Copperwheat} {et~al.}(2010){Copperwheat}, {Marsh}, {Dhillon}, {Littlefair}, {Hickman}, {G{\"a}nsicke}, \& {Southworth}}]{copperwheat-2010}
{Copperwheat}, C.~M., {Marsh}, T.~R., {Dhillon}, V.~S., {et~al.} 2010, \mnras, 402, 1824

\bibitem[{{Covington} {et~al.}(2022){Covington}, {Shaw}, {Mukai}, {Littlefield}, {Heinke}, {Plotkin}, {Barrett}, {Boardman}, {Boyd}, {Brincat}, {Carstens}, {Collins}, {Cook}, {Cooney}, {Fern{\'a}ndez}, {Dufoer}, {Dvorak}, {Galdies}, {Goff}, {Hambsch}, {Johnston}, {Jones}, {Menzies}, {Monard}, {Morelle}, {Nelson}, {{\"O}gmen}, {Rock}, {Sabo}, {Seargeant}, {Stone}, {Ulowetz}, \& {Vanmunster}}]{covington-22}
{Covington}, A.~E., {Shaw}, A.~W., {Mukai}, K., {et~al.} 2022, \apj, 928, 164

\bibitem[{{Cropper}(1990)}]{cropper-90}
{Cropper}, M. 1990, \ssr, 54, 195

\bibitem[{{Cropper} {et~al.}(1998){Cropper}, {Ramsay}, \& {Wu}}]{cropper-98}
{Cropper}, M., {Ramsay}, G., \& {Wu}, K. 1998, \mnras, 293, 222

\bibitem[{{Cutri} {et~al.}(2003){Cutri}, {Skrutskie}, {van Dyk}, {Beichman}, {Carpenter}, {Chester}, {Cambresy}, {Evans}, {Fowler}, {Gizis}, {Howard}, {Huchra}, {Jarrett}, {Kopan}, {Kirkpatrick}, {Light}, {Marsh}, {McCallon}, {Schneider}, {Stiening}, {Sykes}, {Weinberg}, {Wheaton}, {Wheelock}, \& {Zacarias}}]{2mass}
{Cutri}, R.~M., {Skrutskie}, M.~F., {van Dyk}, S., {et~al.} 2003, VizieR Online Data Catalog, II/246

\bibitem[{{Friend} {et~al.}(1990){Friend}, {Martin}, {Smith}, \& {Jones}}]{friend-90}
{Friend}, M.~T., {Martin}, J.~S., {Smith}, R.~C., \& {Jones}, D.~H.~P. 1990, \mnras, 246, 637

\bibitem[{{Garz{\'o}n} {et~al.}(2022){Garz{\'o}n}, {Balcells}, {Gallego}, {Gry}, {Guzm{\'a}n}, {Hammersley}, {Herrero}, {Mu{\~n}oz-Tu{\~n}{\'o}n}, {Pell{\'o}}, {Prieto}, {Bourrec}, {Cabello}, {Cardiel}, {Gonz{\'a}lez-Fern{\'a}ndez}, {Laporte}, {Milliard}, {Pascual}, {Patrick}, {Patr{\'o}n}, {Ram{\'\i}rez-Alegr{\'\i}a}, \& {Streblyanska}}]{garzon-22}
{Garz{\'o}n}, F., {Balcells}, M., {Gallego}, J., {et~al.} 2022, \aap, 667, A107

\bibitem[{{Garz{\'o}n} \& {EMIR Team}(2016)}]{garzon-16}
{Garz{\'o}n}, F. \& {EMIR Team}. 2016, in Astronomical Society of the Pacific Conference Series, Vol. 507, Multi-Object Spectroscopy in the Next Decade: Big Questions, Large Surveys, and Wide Fields, ed. I.~{Skillen}, M.~{Balcells}, \& S.~{Trager}, 297

\bibitem[{{Gray}(1992)}]{libro-gray}
{Gray}, D.~F. 1992, {The Observation and Analysis of Stellar Photospheres} (Cambridge: Cambridge University Press)

\bibitem[{{Harrison}(2016)}]{harrison-16}
{Harrison}, T.~E. 2016, \apj, 833, 14

\bibitem[{{Harrison}(2018)}]{harrison-2018}
{Harrison}, T.~E. 2018, \apj, 861, 102

\bibitem[{{Haswell} {et~al.}(1997){Haswell}, {Patterson}, {Thorstensen}, {Hellier}, \& {Skillman}}]{haswell-97}
{Haswell}, C.~A., {Patterson}, J., {Thorstensen}, J.~R., {Hellier}, C., \& {Skillman}, D.~R. 1997, \apj, 476, 847

\bibitem[{{Hauschildt} {et~al.}(1999){Hauschildt}, {Allard}, \& {Baron}}]{hauschildt-99}
{Hauschildt}, P.~H., {Allard}, F., \& {Baron}, E. 1999, \apj, 512, 377

\bibitem[{{Henden} \& {Honeycutt}(1995)}]{hh-95}
{Henden}, A.~A. \& {Honeycutt}, R.~K. 1995, \pasp, 107, 324

\bibitem[{{Hill} {et~al.}(2022){Hill}, {Littlefield}, {Garnavich}, {Scaringi}, {Szkody}, {Mason}, {Kennedy}, {Shaw}, \& {Covington}}]{katherine-22}
{Hill}, K.~L., {Littlefield}, C., {Garnavich}, P., {et~al.} 2022, \aj, 163, 246

\bibitem[{{Husser} {et~al.}(2013){Husser}, {Wende-von Berg}, {Dreizler}, {Homeier}, {Reiners}, {Barman}, \& {Hauschildt}}]{husser-2013}
{Husser}, T.~O., {Wende-von Berg}, S., {Dreizler}, S., {et~al.} 2013, \aap, 553, A6

\bibitem[{{Joshi}(2012)}]{joshi-2012}
{Joshi}, V. 2012, PhD thesis, Gujarat University, India

\bibitem[{{Kepler} {et~al.}(2016){Kepler}, {Pelisoli}, {Koester}, {Ourique}, {Romero}, {Reindl}, {Kleinman}, {Eisenstein}, {Valois}, \& {Amaral}}]{kepler-2016}
{Kepler}, S.~O., {Pelisoli}, I., {Koester}, D., {et~al.} 2016, \mnras, 455, 3413

\bibitem[{{Kepler} {et~al.}(2019){Kepler}, {Pelisoli}, {Koester}, {Reindl}, {Geier}, {Romero}, {Ourique}, {Oliveira}, \& {Amaral}}]{kepler-2019}
{Kepler}, S.~O., {Pelisoli}, I., {Koester}, D., {et~al.} 2019, \mnras, 486, 2169

\bibitem[{{Knigge}(2006)}]{knigge-2006}
{Knigge}, C. 2006, \mnras, 373, 484

\bibitem[{{Knigge} {et~al.}(2011){Knigge}, {Baraffe}, \& {Patterson}}]{knigge-2011}
{Knigge}, C., {Baraffe}, I., \& {Patterson}, J. 2011, \apjs, 194, 28

\bibitem[{{Maiolino} {et~al.}(2021){Maiolino}, {Titarchuk}, {Wang}, {Frontera}, \& {Orlandini}}]{maiolino-21}
{Maiolino}, T., {Titarchuk}, L., {Wang}, W., {Frontera}, F., \& {Orlandini}, M. 2021, \apj, 911, 80

\bibitem[{{Marsh}(1988)}]{marsh-88}
{Marsh}, T.~R. 1988, \mnras, 231, 1117

\bibitem[{{Marsh} {et~al.}(1994){Marsh}, {Robinson}, \& {Wood}}]{marsh-94}
{Marsh}, T.~R., {Robinson}, E.~L., \& {Wood}, J.~H. 1994, \mnras, 266, 137

\bibitem[{{Martin} {et~al.}(1989){Martin}, {Friend}, {Smith}, \& {Jones}}]{martin-89}
{Martin}, J.~S., {Friend}, M.~T., {Smith}, R.~C., \& {Jones}, D.~H.~P. 1989, \mnras, 240, 519

\bibitem[{{Mateo} {et~al.}(1991){Mateo}, {Szkody}, \& {Garnavich}}]{mateo-91}
{Mateo}, M., {Szkody}, P., \& {Garnavich}, P. 1991, \apj, 370, 370

\bibitem[{{Mukai} {et~al.}(1990){Mukai}, {Mason}, {Howell}, {Allington-Smith}, {Callanan}, {Charles}, {Hassall}, {Machin}, {Naylor}, {Smale}, \& {van Paradijs}}]{mukai-90}
{Mukai}, K., {Mason}, K.~O., {Howell}, S.~B., {et~al.} 1990, \mnras, 245, 385

\bibitem[{{Pala} {et~al.}(2022){Pala}, {G{\"a}nsicke}, {Belloni}, {Parsons}, {Marsh}, {Schreiber}, {Breedt}, {Knigge}, {Sion}, {Szkody}, {Townsley}, {Bildsten}, {Boyd}, {Cook}, {De Martino}, {Godon}, {Kafka}, {Kouprianov}, {Long}, {Monard}, {Myers}, {Nelson}, {Nogami}, {Oksanen}, {Pickard}, {Poyner}, {Reichart}, {Rodriguez Perez}, {Shears}, {Stubbings}, \& {Toloza}}]{pala-2022}
{Pala}, A.~F., {G{\"a}nsicke}, B.~T., {Belloni}, D., {et~al.} 2022, \mnras, 510, 6110

\bibitem[{{Patterson}(1994)}]{patterson-94}
{Patterson}, J. 1994, \pasp, 106, 209

\bibitem[{{Patterson} {et~al.}(2005){Patterson}, {Kemp}, {Harvey}, {Fried}, {Rea}, {Monard}, {Cook}, {Skillman}, {Vanmunster}, {Bolt}, {Armstrong}, {McCormick}, {Krajci}, {Jensen}, {Gunn}, {Butterworth}, {Foote}, {Bos}, {Masi}, \& {Warhurst}}]{patterson-2005}
{Patterson}, J., {Kemp}, J., {Harvey}, D.~A., {et~al.} 2005, \pasp, 117, 1204

\bibitem[{{Patterson} {et~al.}(1982){Patterson}, {Schwartz}, {Bradt}, {Remillard}, {McHardy}, {Pye}, {Williams}, {Fesen}, \& {Szkody}}]{patterson-82}
{Patterson}, J., {Schwartz}, D.~A., {Bradt}, H., {et~al.} 1982, in Bulletin of the American Astronomical Society, Vol.~14, 618

\bibitem[{{Patterson} {et~al.}(1992){Patterson}, {Schwartz}, {Pye}, {Blair}, {Williams}, \& {Caillault}}]{patterson-92}
{Patterson}, J., {Schwartz}, D.~A., {Pye}, J.~P., {et~al.} 1992, \apj, 392, 233

\bibitem[{{Pecaut} \& {Mamajek}(2013)}]{pecaut-13}
{Pecaut}, M.~J. \& {Mamajek}, E.~E. 2013, \apjs, 208, 9

\bibitem[{{Rayner} {et~al.}(2009){Rayner}, {Cushing}, \& {Vacca}}]{rayner-2009}
{Rayner}, J.~T., {Cushing}, M.~C., \& {Vacca}, W.~D. 2009, \apjs, 185, 289

\bibitem[{{Robinson} {et~al.}(1986){Robinson}, {Zhang}, \& {Stover}}]{robinson-86}
{Robinson}, E.~L., {Zhang}, E.~H., \& {Stover}, R.~J. 1986, \apj, 305, 732

\bibitem[{{Rodr{\'\i}guez-Gil} {et~al.}(2015){Rodr{\'\i}guez-Gil}, {Shahbaz}, {Marsh}, {G{\"a}nsicke}, {Steeghs}, {Long}, {Mart{\'\i}nez-Pais}, {Armas Padilla}, {Schwarz}, {Schreiber}, {Torres}, {Koester}, {Dhillon}, {Castellano}, \& {Rodr{\'\i}guez}}]{pablo-2015}
{Rodr{\'\i}guez-Gil}, P., {Shahbaz}, T., {Marsh}, T.~R., {et~al.} 2015, \mnras, 452, 146

\bibitem[{{Rodr{\'\i}guez-Gil} {et~al.}(2020){Rodr{\'\i}guez-Gil}, {Shahbaz}, {Torres}, {G{\"a}nsicke}, {Izquierdo}, {Toloza}, {{\'A}lvarez-Hern{\'a}ndez}, {Steeghs}, {van Spaandonk}, {Koester}, \& {Rodr{\'\i}guez}}]{pablo-2020}
{Rodr{\'\i}guez-Gil}, P., {Shahbaz}, T., {Torres}, M.~A.~P., {et~al.} 2020, \mnras, 494, 425

\bibitem[{{Rojas-Ayala} {et~al.}(2012){Rojas-Ayala}, {Covey}, {Muirhead}, \& {Lloyd}}]{rojas-ayala-2012}
{Rojas-Ayala}, B., {Covey}, K.~R., {Muirhead}, P.~S., \& {Lloyd}, J.~P. 2012, \apj, 748, 93

\bibitem[{{Shahbaz} {et~al.}(2000){Shahbaz}, {Groot}, {Phillips}, {Casares}, {Charles}, \& {van Paradijs}}]{tariq-2000}
{Shahbaz}, T., {Groot}, P., {Phillips}, S.~N., {et~al.} 2000, \mnras, 314, 747

\bibitem[{{Shahbaz} {et~al.}(2017){Shahbaz}, {Linares}, \& {Breton}}]{tariq-2017}
{Shahbaz}, T., {Linares}, M., \& {Breton}, R.~P. 2017, \mnras, 472, 4287

\bibitem[{{Shahbaz} {et~al.}(2019){Shahbaz}, {Linares}, {Rodr{\'\i}guez-Gil}, \& {Casares}}]{shahbaz-2019}
{Shahbaz}, T., {Linares}, M., {Rodr{\'\i}guez-Gil}, P., \& {Casares}, J. 2019, \mnras, 488, 198

\bibitem[{{Shahbaz} {et~al.}(2014){Shahbaz}, {Watson}, \& {Dhillon}}]{tariq-14}
{Shahbaz}, T., {Watson}, C.~A., \& {Dhillon}, V.~S. 2014, \mnras, 440, 504

\bibitem[{{Shahbaz} {et~al.}(2003){Shahbaz}, {Zurita}, {Casares}, {Dubus}, {Charles}, {Wagner}, \& {Ryan}}]{tariq-2003}
{Shahbaz}, T., {Zurita}, C., {Casares}, J., {et~al.} 2003, \apj, 585, 443

\bibitem[{{Shaw} {et~al.}(2020){Shaw}, {Heinke}, {Mukai}, {Tomsick}, {Doroshenko}, {Suleimanov}, {Buisson}, {Gand hi}, {Grefenstette}, {Hare}, {Jiang}, {Ludlam}, {Rana}, \& {Sivakoff}}]{shaw-2020}
{Shaw}, A.~W., {Heinke}, C.~O., {Mukai}, K., {et~al.} 2020, \mnras, 498, 3457

\bibitem[{{Smette} {et~al.}(2015){Smette}, {Sana}, {Noll}, {Horst}, {Kausch}, {Kimeswenger}, {Barden}, {Szyszka}, {Jones}, {Gallenne}, {Vinther}, {Ballester}, \& {Taylor}}]{smette-15}
{Smette}, A., {Sana}, H., {Noll}, S., {et~al.} 2015, \aap, 576, A77

\bibitem[{{Steeghs} \& {Jonker}(2007)}]{steeghs-2007}
{Steeghs}, D. \& {Jonker}, P.~G. 2007, \apjl, 669, L85

\bibitem[{{Suleimanov} {et~al.}(2005){Suleimanov}, {Revnivtsev}, \& {Ritter}}]{suleimanov-2005}
{Suleimanov}, V., {Revnivtsev}, M., \& {Ritter}, H. 2005, \aap, 435, 191

\bibitem[{{Suleimanov} {et~al.}(2019){Suleimanov}, {Doroshenko}, \& {Werner}}]{suleimanov-19}
{Suleimanov}, V.~F., {Doroshenko}, V., \& {Werner}, K. 2019, \mnras, 482, 3622

\bibitem[{{Szkody} {et~al.}(2002){Szkody}, {Nishikida}, {Erb}, {Mukai}, {Hellier}, {Uemura}, {Kato}, {Pavlenko}, {Katysheva}, {Shugarov}, \& {Cook}}]{szkody-2002}
{Szkody}, P., {Nishikida}, K., {Erb}, D., {et~al.} 2002, \aj, 123, 413

\bibitem[{{Torres} {et~al.}(2010){Torres}, {Andersen}, \& {Gim{\'e}nez}}]{guilletorres-2010}
{Torres}, G., {Andersen}, J., \& {Gim{\'e}nez}, A. 2010, \aapr, 18, 67

\bibitem[{{Torres} {et~al.}(2020){Torres}, {Casares}, {Jim{\'e}nez-Ibarra}, {{\'A}lvarez-Hern{\'a}ndez}, {Mu{\~n}oz-Darias}, {Armas Padilla}, {Jonker}, \& {Heida}}]{torres-2020}
{Torres}, M.~A.~P., {Casares}, J., {Jim{\'e}nez-Ibarra}, F., {et~al.} 2020, \apjl, 893, L37

\bibitem[{{{\v{S}}imon}(2000)}]{simon-2000}
{{\v{S}}imon}, V. 2000, \aap, 360, 627

\bibitem[{{Warner}(1995)}]{warner-libro}
{Warner}, B. 1995, Cambridge Astrophysics Series, 28

\bibitem[{{Williams}(1983)}]{williams-83}
{Williams}, G. 1983, \apjs, 53, 523

\bibitem[{{Wilson} {et~al.}(2020){Wilson}, {Devinney}, \& {Van Hamme}}]{wilson-2020}
{Wilson}, R.~E., {Devinney}, E.~J., \& {Van Hamme}, W. 2020, {WD: Wilson-Devinney binary star modeling}, Astrophysics Source Code Library, record ascl:2004.004

\bibitem[{{Wu} {et~al.}(1994){Wu}, {Chanmugam}, \& {Shaviv}}]{wu-94}
{Wu}, K., {Chanmugam}, G., \& {Shaviv}, G. 1994, \apj, 426, 664

\bibitem[{{Xu} {et~al.}(2019){Xu}, {Yu}, \& {Li}}]{xu-2019}
{Xu}, X.-j., {Yu}, Z.-l., \& {Li}, X.-d. 2019, \apj, 878, 53

\bibitem[{{Yuasa} {et~al.}(2010){Yuasa}, {Nakazawa}, {Makishima}, {Saitou}, {Ishida}, {Ebisawa}, {Mori}, \& {Yamada}}]{yuasa-2010}
{Yuasa}, T., {Nakazawa}, K., {Makishima}, K., {et~al.} 2010, \aap, 520, A25

\bibitem[{{Zorotovic} {et~al.}(2011){Zorotovic}, {Schreiber}, \& {G{\"a}nsicke}}]{zorotovic-2011}
{Zorotovic}, M., {Schreiber}, M.~R., \& {G{\"a}nsicke}, B.~T. 2011, \aap, 536, A42

\end{thebibliography}


\end{document}